\documentclass[useAMS,usenatbib,letter]{mn2e}
\usepackage{graphicx,lscape}

\def\etal{et~al. }

\def\aap{A \&\ A}

\def\aj{AJ}
\def\apj{ApJ}
\def\apjl{ApJL}

\def\araa{ARA \&\ A}

\def\mnras{MNRAS}
\def\pasj{PASJ}

\title[Effect of Stellar Galactic Environments on Planetary Discs
I. The Sun] {Effect of Different Stellar Galactic Environments on
Planetary Discs I: The Solar Neighbourhood and the Birth Cloud of the
Sun.}

\author[J.J. Jim\'enez-Torres, B. Pichardo, G. Lake, H. Throop]{Juan J. Jimenez-Torres$^{1}$,
Barbara Pichardo$^{1}$\thanks{E-mail: barbara@astroscu.unam.mx (B.P.)}, George Lake$^{2}$, Henry Throop$^{1,3}$\\
$^{1}$Instituto de Astronom\'\i a, Universidad Nacional Aut\'onoma de M\'exico, Apdo. postal 70-264 Ciudad Universitaria, D.F., M\'exico \\ $^{2}$Institute for Theoretical Physics, University of Z\"urich, CH-8057 Z\"urich, Switzerland \\ $^{3}$Southwest Research Institute, Department of Space Studies, 1050 Walnut St, Ste 300, Boulder, CO 80302, USA}

\begin{document}

\date{Accepted . Received ; in original form }

\pagerange{\pageref{firstpage}--\pageref{lastpage}} \pubyear{2011}

\maketitle

\label{firstpage}

\begin{abstract} 
  We have computed trajectories, distances and times of closest
  approaches to the Sun by stars in the Solar neighborhood with known
  position, radial velocity and proper motions. For this purpose we
  have used a full potential model of the Galaxy that reproduces the
  local z-force, the Oort constants, the local escape velocity, and
  the rotation curve of the Galaxy. From our sample we constructed
  initial conditions, within observational uncertainties, with a Monte
  Carlo scheme for the twelve most suspicious candidates because of
  their small tangential motion. We find that the star Gliese 710 will
  have the closest approach to the Sun, with a distance of
  approximately 0.34 pc at 1.36 Myr in the future. We show that the
  effect of a flyby with the characteristics of Gliese 710 on a 100 AU
  test particle disk representing the Solar system is
  negligible. However, since there is a lack of 6D data for a large
  percentage of stars in the Solar neighborhood, closer approaches may
  exist.  We calculate parameters of passing stars that would cause
  noticeable effects on the Solar disk. Regarding the birth cloud of
  the Sun, we performed experiments to reproduce roughly the observed
  orbital parameters such as eccentricities and inclinations of the
  Kuiper Belt. It is known now that in Galactic environments, such as
  stellar formation regions, the stellar densities of new born stars,
  are high enough to produce close encounters within 200 AU. Moreover,
  in these Galactic environments, the velocity dispersion is
  relatively low, typically $\sigma\sim$ 1-3 km s$^{-1}$. We find that
  with a velocity dispersion of $\sim$1 km s$^{-1}$ and an approach
  distance of about 150 AU, typical of these regions, we obtain
  approximately the eccentricities and inclinations seen in the
  current Solar system. Simple analytical calculations of stellar
  encounters effects on the Oort cloud are presented.
\end{abstract}


\section{Introduction}\label{Intro}
Observations of extrasolar planets (e.g. Schneider 2010) show that
planetary orbits in other planetary systems are disordered, showing a
wide range of eccentricities. For extrasolar planets with semi-major
axes $a \ge 0.1$ AU, the mean of the eccentricity distribution is
$e\approx 0.3$ and the median is $e\approx 0.24$ (Adams 2010). Thus,
available data indicate that planetary systems discovered the last
decade are more dynamically active and disordered than our own (Udry
\& Santos 2007; Adams 2010).

A study of planetary disc dynamics under the stellar influence of
different Galactic environments is presented in a set of papers. In
this work we introduce the first two Galactic environments related to
the Sun: the Solar neighborhood and the birth cloud of the Sun.

The Solar neighborhood has been defined as the region of space
centered on the Sun that is much smaller than the overall size of the
Galaxy, and whose contents are known with reasonable completeness
(Gilmore 1992; Binney \& Tremaine 2007). Now, from a theoretical
simple approximation, it seems clear that the probability in the
current Solar neighborhood to have an important encounter for the
planetary system, with another star (i.e. less than 300 AU for a solar
mass stellar flyby), is almost negligible. Let us for example,
consider a typical stellar density for the Solar neighborhood of
approximately 0.05 M$_\odot$ pc$^3$, we can calculate the mean free
path, $\lambda$, for approaches within say 300 AU, as the radius of a
cross section, $\sigma$, and setting 50 km/s as the typical velocity
dispersion, $\sigma_v$, in the Solar neighborhood, we find that the
time necessary, in the current conditions of velocity dispersion and
density of the Solar neighborhood, to see an encounter within 300 AU
with the Sun (or between any couple of stars in these dynamical
conditions), would be, $t\approx\lambda/\langle v\rangle\approx
1/\sigma n \sigma_v$, that corresponds to approximately three Hubble
times. On the other hand, Garc\'\i a-S\'anchez \etal (2001), by
comparing Hipparcos observations with the stellar luminosity function
for star systems within 50 pc of the Sun, estimate that only about
one-fifth of the star systems were detected by Hipparcos, and they
correct for that incompleteness in the data obtaining about 12 stellar
encounters per Myr within one pc of the Sun.

However, in the case of the Solar system it is clear that a rough
approximation, is not good enough. First with Hipparcos, and some
knowledge of the local and global Galactic potential we are now able
to compute orbits of neighboring stars and determinate at good
approximation distances, times, etc. to stars in the Solar
neighborhood at every time within a few million years. This will
improve enormously in the near future with the advent of large surveys
of the Milky Way such as GAIA. For now we are not close to having a
complete set of 6-dimensional information (position and velocity) of
all stars near the Sun. In this paper we take the nearest stars to the
Sun with proper motions, parallaxes and radial velocities and we
compute their past and future trajectories as well as their closest
approach, distance and time. Thus, for our purposes, the Solar
neighborhood is the radius of the sphere that contains all the stars
(with 6-dimensional parameters known), whose maximum approach to the
Sun was or will be less than 5 pc within 10 Myr to the past or to the
future, this is approximately 200 pc.

On the other hand, even when, in the current Solar neighborhood
conditions, the probability of close stellar approaches is almost
negligible, it is now known that as many as 90\% of stars appear to
form in clusters or groups with $10^2$ to $10^3$ members (Carpenter
2000; Lada \& Lada 2003; Clark et al. 2005; Adams 2010). Short-lived
radioisotopes in Solar meteorites argue that formation happened near
at least one massive star, probably in a large cluster (Goswami \&
Vanhala 2000; Meyer \& Clayton 2000; Hester et al. 2004; Looney et
al. 2006; Wadhwa et al. 2007).  In their early stages, most stars were
in relatively crowded environments. In such environments, close
stellar encounters would be frequent and affect the stability of
planetary systems around the stars (de la Fuente Marcos \& de la
Fuente Marcos 1997; Laughlin \& Adams 1998; Hurley \& Shara 2002;
Pfahl \& Muterspaugh 2006; Spurzem \etal 2010).

Some work has been devoted to searching for stellar perturbers of the
cometary disks and clouds. Mathews (1994) identified close approaches
for six stars within the next $5 \times 10^{4}$ yr, within a radius of
about 5 pc. Garc\'\i a-S\'anchez \etal (1997) started a search for
stars passing close to the Sun using Hipparcos data, assuming a
straight line motion model. Subsequently Garc\'\i a-S\'anchez \etal
(1999) continued their search by integrating the motion of the
candidate stars and the Sun in a local Galactic potential and in a
simple global Galactic Potential. More recently, Bobylev (2010)
presents a study of the closest encounters of stars in the Solar
neighborhood with the Sun, using also a simple local approximation to
integrate orbits.

Of course, a star passing near the Sun has greater effect on the Oort
cloud -- an extremely extended structure -- than on the planetary
disk. However, in the solar system, the Kuiper belt has properties
acquired early in its history, and some of these are difficult to
explain under the assumption that the solar system has been always
isolated, such as the excitation of the eccentricities and
inclinations in the classic belt; the mass deficit of the Kuiper belt,
and the sharp outer edge of the classical belt at approximately 48
AU. Beyond this boundary, only high-eccentricity objects typical of
the scattered disk or of the detached population seem to exist, and
finally, the existence of some mysterious large bodies (approximately
Pluto size), with extreme eccentricities and perihelia (Sedna). One of
the most accepted theories to explain the characteristics of Kuiper
Belt, is then based on the idea that a close stellar passage could have
taken place in the early history of our planetary system (Ida et
al. 2000; Kobayashi \& Ida 2001). In this paper we experiment with
stellar encounters on a 100 AU particle disk, looking for the minimum
distance where an encounter is important, and we seek to produce
Kuiper belt orbital characteristics such as eccentricities and
inclinations.

This paper is organized as follows. In Section \ref{method}, we
describe the methods and numerical implementation. In Section
\ref{results} we present the stellar sample of the Solar neighborhood
and our results for this first Galactic environment. In Section
\ref{sun_birth_cloud_simulations} we show the results of the Sun's
birth cluster. Conclusions are presented in Section \ref{conclusions}.

\section{Methodology and Numerical Implementation}\label{method}
We have implemented two codes to solve the equations of motion. The
first calculates stellar trajectories in the Solar vicinity with all
their observational orbital parameters known (positions and
velocities) in a Milky-Way-like potential. The second simulates debris
discs under the influence of stellar encounters.

\subsection{The Solar Neighborhood Code}\label{SNC}
We calculate the trajectory of stars with known 6D data (derived from
$\alpha$, $\delta$, radial velocity and proper motions known) to
determinate the distance and time (past or future) of their closest
approach to the Sun.

Instead of using a straight line approximation to solve the stellar
orbits in the Solar neighborhood, we solved the orbits in a
Milky-Way-like Galactic potential (Pichardo et al. 2003, 2004).

The axisymmetric part of the model consists of a background potential
with a bulge, a flattened disk (Miyamoto \& Nagai 1975) with a
scale-height of 250 pc, and a massive halo extending to a radius of
100 kpc. The model has a total mass of $9\times 10^{11}$ M$_{\odot}$,
with a local escape velocity of 536 km s$^{-1}$. This potential
satisfies observational constraints such as the Galactic rotation
curve, with a rotation velocity of 220 km s$^{-1}$ at $R_0 = 8.5$ kpc,
the perpendicular force at the Solar circle, and the Oort constants,
among others (for more details see Pichardo et al. 2003, 2004; Martos
et al. 2004; Antoja et al. 2009).

Given data in the equatorial system are transformed to Galactic
coordinates. The equations of motion are solved with a Bulirsh-Stoer
adaptive integrator (Press \etal 1992) that gives relative errors for
the integrals of motion (total energy, z component of angular
momentum, for the axisymmetric potential, or Jacobi constant, for the
barred or armed potential) of 10$^{-10}$ in the worst case (when the
non axisymmetric components of the Galaxy are included). The Sun is
located 8.5 kpc from the Galactic nucleus and 0.035 kpc above the the
plane of the disc. The velocity along the $x$ axis is -9 km s$^{-1}$
and on the $y$ axis is -220 km s$^{-1}$.

This code computes distances, times, velocities and uncertainties for
stars at the moment of maximum approach to the Sun within
observational uncertainties. These are obtained from catalogs, and
data basis mainly (Simbad, Hipparcos, Nexxus 2) and papers (Bower
\etal 2009; Dybczynski 2006; Garc\'\i a-S\'anchez \etal 1999). Initial
conditions are constructed as normal random values within
observational uncertainties (note that it is the uncertainty in
parallax the one with the normal distribution, not the uncertainty in
distance). Ten thousand orbits were computed as initial conditions for
each star in order to obtain the final error bars.

In the original version of the code, additionally to the background
axisymmetric potential, it included non-axisymmetric features: spiral
arms and bar (Pichardo \etal 2003, 2004). In the same way as Garc\'\i
a-S\'anchez \etal 2001, who constructed a potential including a very
simple axisymmetric background and weak spiral arms (pitch angle of
$\sim6.9^o$), we conclude that the difference between using a full
Galactic model and an axisymmetric model among the Solar neighborhood
stars is negligible (within $\approx 3\%$ for the most distant stars
to the Sun in this study). Although there is evidence that the
existence of spiral arms and bars have direct influence on the local
stellar distribution and evolution (Dehnen 2000; Chakrabarty 2007;
Antoja \etal 2009), the time we run the orbits is rather short ($\leq
10^7$ years), much less than a dynamical time, and not enough to
produce a noticeable difference in the orbits produced by the
non-axisymmetric potential. To facilitate the calculations we
performed the final calculations only with the axisymmetric background
potential of the Galaxy, since the bar and arms are not of importance
in this time scale.

\subsection{The Stellar Encounter Code}\label{SEC}

The second code is a 3D particle code that calculates the interaction
of a planetary system (a central star and a disk of test particles)
with a second stellar body (a flyby star).

The sampling of test particles goes as $a\propto n^{-3/2}$, where $a$
is the initial radius of a particle orbit and $n$ is the number of the
orbit; these test particles are affected by the gravitational forces
of both the central star and the flyby star; and the equations of
motion are solved for the central star’s (Sun’s) non-inertial frame of
reference. The code calculates the main orbital characteristics of
each of the test particles simulating a debris disk after the flyby;
it calculates the semi-major axis, the eccentricity, and the
inclinations. The Bulirsh-Stoer integrator gives a maximum relative
error before the flyby of $10^{-14}$ and $10^{-13}$ in the energy and
angular momentum integrals, respectively.

\begin{figure}
\includegraphics[width=84mm]{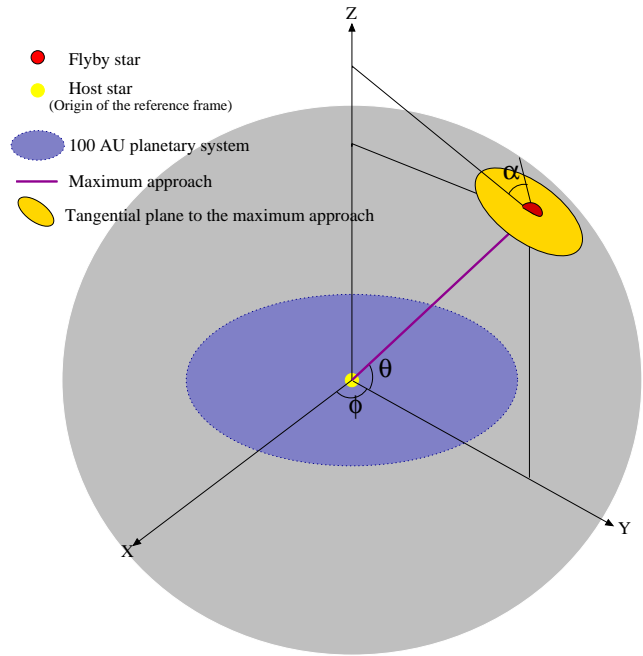}
\vspace{0.1cm}
\caption {Schematic figure of a stellar encounter on a planetary
  system by a flyby star. The disc is simulated with test
  particles. The initial conditions are particles on circular orbits
  with zero inclination.}
\label{fig.scheme}
\end{figure}

In Figure \ref{fig.scheme}, we show the schematics of the relevant
parameters used on the code of a stellar encounter. The dark disk in
the center of the system represents the planetary disk, the gray
sphere represents the radius of the minimum distance of the flyby, and
the bright disk is tangent to the sphere at the point of minimum
distance.  The flyby attack angles are: $\phi$, the
azimuthal angle with respect to the disc, it goes from 0$^o$ to
360$^o$; $\theta$, the polar angle with respect to the disc,
goes from -90$^o$ to 90$^o$; and $\alpha$, the angle between
the flyby plane orbit and the symmetry axis of the planetary disc, it
goes from 0$^o$ to 360$^o$.

With the purpose of covering several Galactic stellar environments,
and to acquire a physical idea of what parameters are the most
relevant in this kind of interaction, besides the specific
applications of this code to the Sun's birth cluster and the current
Solar environment, we computed more than a thosand experiments
covering the different Galactic ranges in distance (stellar closest
approximation), velocities, angles of flyby trajectories and stellar
masses. For distance: 5 to 1000 AU, for velocity: 1 to 180 km/s, for
mass: 1 Jupiter mass to 4 Solar masses. Regarding flyby attack angle
$\phi$, due to the symmetry of the problem, this entrance angle is
indistinct; we took for our experiments $\phi=0^o$.

In the application presented in this paper the effects of the exact
direction of entrance of the flyby resulted of no importance, this,
due to the fact that: a) in the current Solar neighborhood stars will
not come close enough for the angle to be important and the effect of
encounters at such large distances would be negligible, even for
objects in the Oort cloud in the majority of cases; b) in the case of
the birth cloud of the Sun, stars approaches might have been
important, however in this case the orientation of discs with respect
to flybys are unknown and probably rather random.

Since we are talking about one single encounter in this Galactic
environment, we chose a general interaction, this is 45$^o$ for
$\theta$ and $\alpha$, that should produce an intermediate effect
produced by all possible inclinations of the orbital plane. This
means, the flyby orbit will enter at 45$^o$ with respect to the plane
of the disc, and 45$^o$ with respect to the disc axis. It is worth to
mention, that the results are nearly insensitive to changes of
$\alpha$, while changes in $\theta$ produce diferent results being
larger for $\theta=0^o$. A global study of the parameters will be
presented on a future paper.

\section{RESULTS}\label{results}
Here we present the first two Galactic regions studied in
this work.

\subsection{The Solar Neighborhood}\label{secSN}
For the Solar neighborhood, we looked for the closest stars to the Sun
with known position, radial velocity and proper motions. It is worth
to mention that for the nature of the sample, this is, we are able
only to compute stars with 6D known information from literature, our
sample is biased to the closest, more massive and with the largest
proper motion stars. However, this bias actually represents the most
important candidates to perturb our planetary disc. Considering this
bias, from the more than a thousand stars in the full sample, we find
that 67 stars passed or will pass within 3 pc from the Sun, between
the last 8 Myr and the next 5 Myr, this is roughly, 5 stars per
Myr. Only one of this 67 set, will barely perturb the outer Oort
cloud, and the planetary disc will not notice that star in the
dynamical sense. On the other hand, it should be noticed that there
could be a star of which we do not have 6D information that could come
even closer than G710. Even in our sample, we could have stars that
will approach to the Sun several times along the stellar trajectories
around the Galaxy, but in this case, there is no model of the Galaxy
good enough to warranty precise orbits for these stars for more than a
few million years.

\subsubsection{The Stellar Sample}\label{stellar_sample}

We used parallax and proper-motion data and uncertainties from the
Hipparcos catalog for 1167 nearby stars suspected to pass close to the
Solar System. We take radial velocity measurements from different
catalogs (Hipparcos, Simbad, Nexxus2) and papers (Bower \etal 2009;
Dybczynski 2006; Garc\'\i a-S\'anchez \etal 1999) to produce
six-dimensional trajectories of the whole sample of stars, and
calculate the distance at the point of maximum approach to the Sun,
the time and velocity (the whole table is available on request via
e-mail). From this sample, we produced a subsample of 67 stars with
their observational uncertainties and calculated closest approach
distances to the Sun, within 3 pc. From this subsample, we ran 34
stars toward the past (those receding now), and 33 stars toward the
future (those approaching now) looking for those passages that might
perturb on the Solar System. In particular, these 67 stars are
potential perturbers of the Oort Cloud (Garc\'\i a-S\'anchez, \etal
1999).

\begin{figure}
\includegraphics[width=90mm]{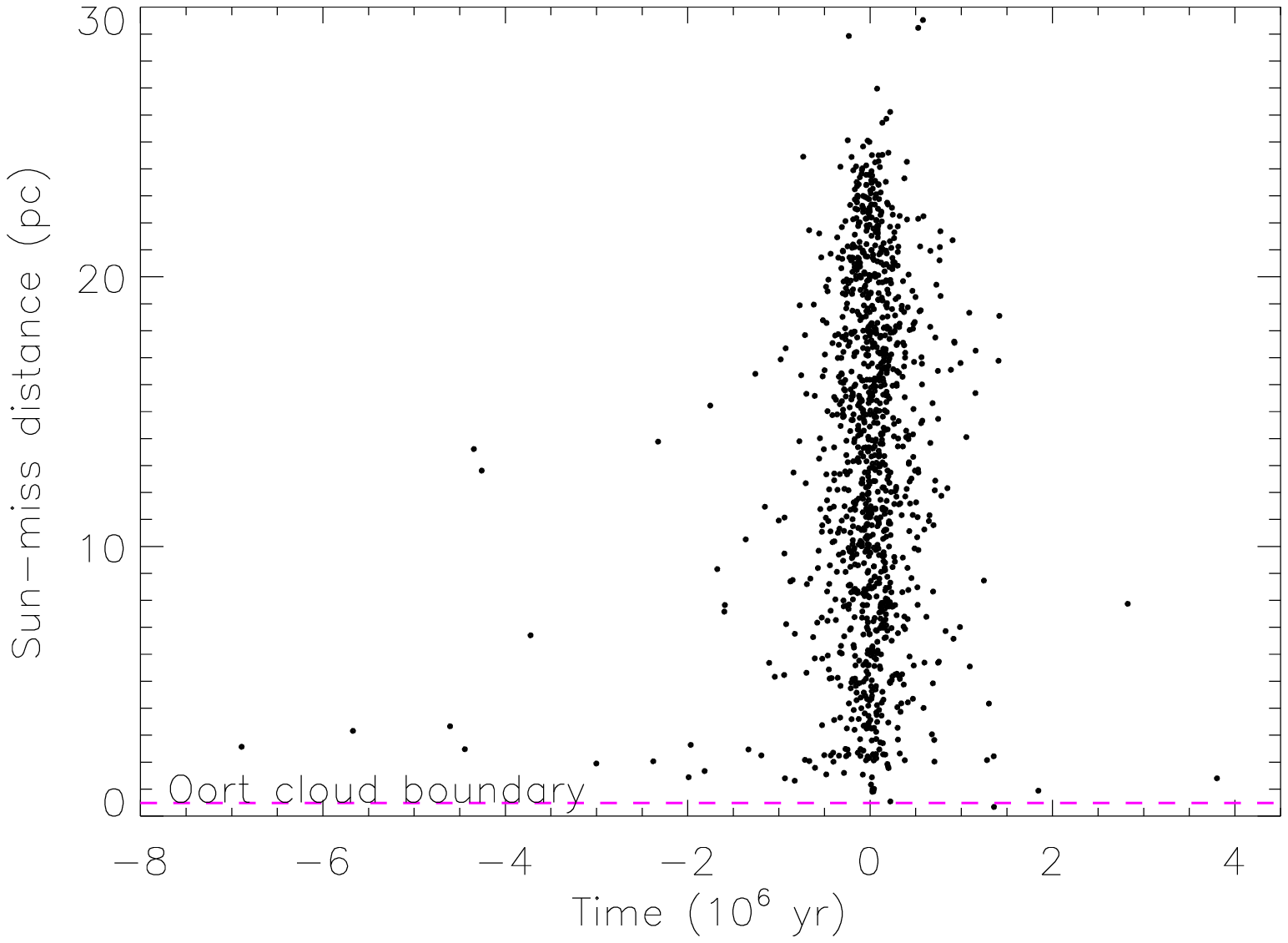}
\includegraphics[width=90mm]{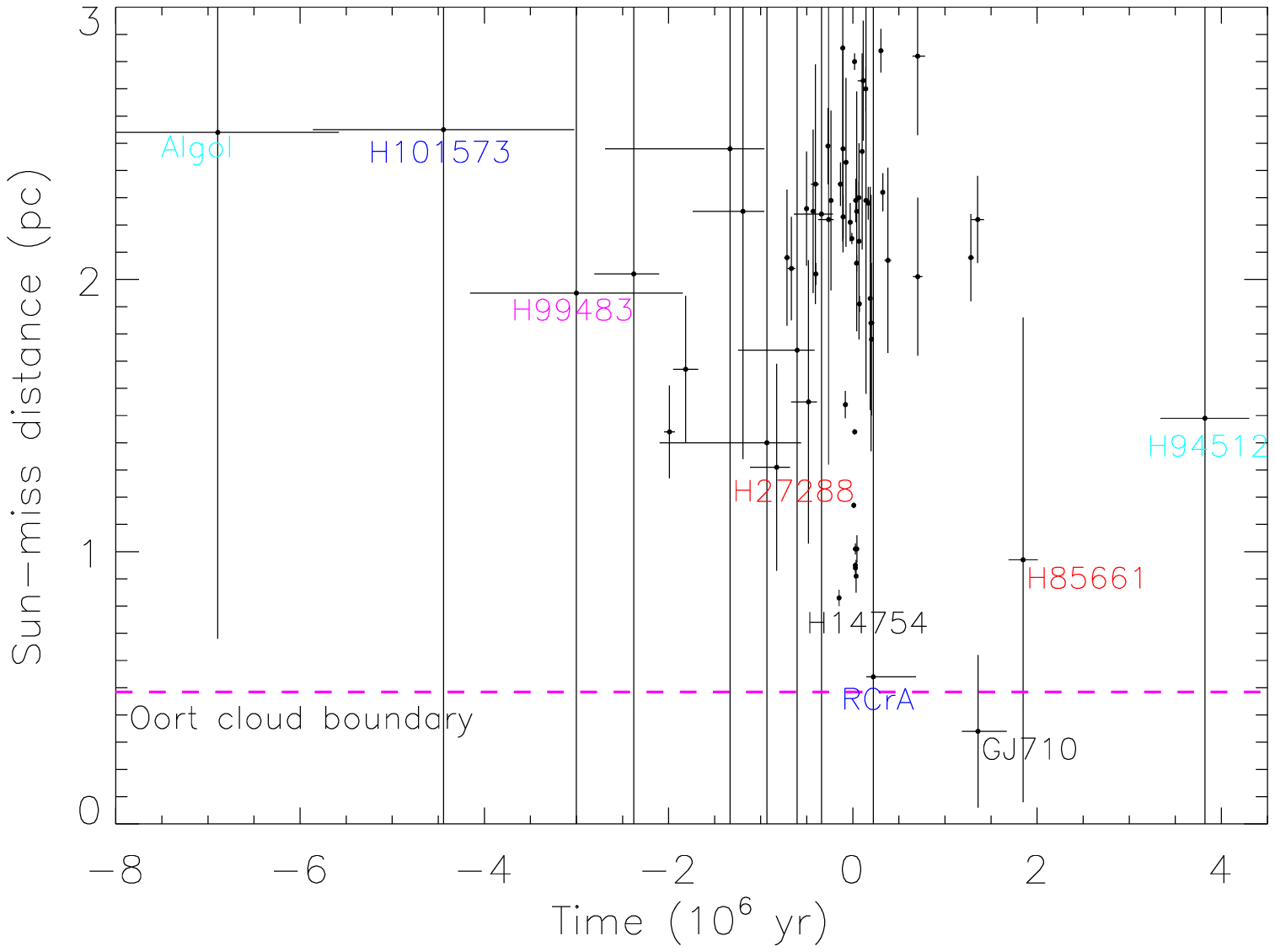}
\vspace{0.1cm}
\caption {Miss distance vs. time of stellar approaches for those stars
in the total sample of 1167 stars within 30 pc as maximum approach
distance to the Sun. The Oort cloud boundary at $\sim$0.48 pc is
plotted as reference. The lower frame is a zoom that shows a subsample
with the closest approach distances (less than 3 pc) to the Sun (see
Tables \ref{table.past} and \ref{table.future}) and their computed
uncertainties. Some star's names are illustrated in the figure.}
\label{fig.oortcloud}
\end{figure}

The stars with the closest approach distances, within 3 pc from the
Sun, are listed in Tables \ref{table.past} and \ref{table.future}
sorted first by receding versus approaching then by parallax. Columns
are arranged as follows, star's name, parallax, radial velocity,
minimum approach distance to the Sun, approach time, and relative
velocity. These predicted passages are contained in a time interval of
about -8 and +4 Myr. The close-approach distances vs. time are shown
in Figure \ref{fig.oortcloud}. In the upper panel of this figure we
show the variation with time of the separation distance between each
star and the Sun, at the moment of maximum approach, for all stars in
our sample (of 1167 stars) which closest distance is less than 30
pc. The lower frame is a zoom that shows the stars with the closest
approach distances, within 3 pc (Tables \ref{table.past} and
\ref{table.future}), both plots for the time interval, from -8 Myr to
4 Myr. Stars coming within 2-3 pc may perturb the Oort cloud. The star
with the closest future approach is Gliese 710 (GJ 710). The predicted
minimum distance for this star is 0.34 pc, in 1.36 Myr. This is the
only star in the sample with an approach distance less than the radius
of the Oort cloud $\sim$ 0.48 pc.

\begin{table*}
\caption{Astronomical data for the 34 stellar subsample run toward the
past.}
\scalebox{0.9}[1.3]{
\begin{tabular}{@{}lr@{}lr@{$\pm$}lr@{$\pm$}lr@{$\pm$}lr@{}lr@{}lr@{}l@{}}
\hline
Star Name        & \multicolumn{2}{c}{Parallax}         & \multicolumn{2}{c}{PM($\alpha$)} & \multicolumn{2}{c}{PM($\delta$)} & \multicolumn{2}{c}{V$_r$}         & \multicolumn{2}{c}{Miss Distance} & \multicolumn{2}{c}{Time}         & \multicolumn{2}{c}{Vel$_{App}$} \\

& \multicolumn{2}{c}{(arcsec)}         & \multicolumn{2}{c}{(arcsec/yr)} & \multicolumn{2}{c}{(arcsec/yr)} & \multicolumn{2}{c}{(km/s)}         & \multicolumn{2}{c}{(pc)} & \multicolumn{2}{c}{(10$^3$ yr)}         & \multicolumn{2}{c}{(km/s)} \\ 
\hline

GJ65B(LHS10)      & 0.3737&$\pm$0.0060 &  3.3210 & 0.0050 &  0.5620 & 0.0050 &  29.0 &  2.0 &  2.21  &   $\pm$0.07      &  -28.5   & $\pm$0.9                    &  51.64 & $\pm$1.38 \\
H16537(GJ144)     & 0.3108&$\pm$0.0009 & -0.9764 & 0.0010 &  0.0180 & 0.0009 &  15.5 &  0.9 &  2.23  &   $\pm$0.08      & -105.5   & $\pm$0.4                    &  21.50 & $\pm$0.73 \\
H5643(GJ54.1)     & 0.2691&$\pm$0.0076 &  1.2101 & 0.0052 &  0.6470 & 0.0039 &  28.0 &  5.0 &  2.43  &   $\pm$0.31      &  -74.4   & ${+3.7}\atop{-0.6}$         & 36.99  & $\pm$4.22 \\
H24186(GJ191)     & 0.2553&$\pm$0.0009 &  6.5061 & 0.0010 & -5.7314 & 0.0009 & 245.5 &  2.0 &  2.15  &   $\pm$0.02      &  -10.9   & $\pm$0.1                    & 293.58 & $\pm$1.90 \\
H30920(GJ234A)    & 0.2429&$\pm$0.0026 &  0.6947 & 0.0030 & -0.6186 & 0.0025 &  24.0 &  5.0 &  2.48  &   $\pm$0.38      & -106.7   & ${+8.4}\atop{-2.8}$         &  30.09 & $\pm$4.41 \\
H103039(LP816-60) & 0.1820&$\pm$0.0037 & -0.3067 & 0.0038 &  0.0308 & 0.0041 &  15.8 &  0.6 &  2.49  &   $\pm$0.14      & -270.2   & $\pm$7.8                    &  17.72 & $\pm$0.61 \\
H33226(GJ251)     & 0.1813&$\pm$0.0019 & -0.7293 & 0.0021 & -0.3993 & 0.0013 &  36.0 & 10.0 &  2.85  &   $\pm$0.71      & -109.8   & ${+15.7}\atop{-8.8}$        &  42.06 & $\pm$9.35 \\
H40501(GJ2066)    & 0.1090&$\pm$0.0018 & -0.3750 & 0.0022 &  0.0601 & 0.0015 &  62.2 &  0.1 &  2.35  &   $\pm$0.08      & -134.7   & $\pm$2.1                    &  64.36 & $\pm$0.13 \\
H14754(HD20523)   & 0.0985&$\pm$0.0015 &  0.0419 & 0.0015 & -0.1041 & 0.0019 &  65.9 &  0.1 &  0.83  &   $\pm$0.03      & -149.6   & $\pm$2.4                    &  66.12 & $\pm$0.11 \\
H22738(GJ2036A)   & 0.0890&$\pm$0.0036 &  0.1329 & 0.0043 &  0.0739 & 0.0038 &  40.1 & 10.0 &  2.22  &   $\pm$0.90      & -263.2   & ${+52.4\hfill}\atop{-112.5}$&  40.92 & $\pm$10.95 \\
H26335(GJ208)     & 0.0879&$\pm$0.0013 & -0.0026 & 0.0013 & -0.0576 & 0.0009 &  22.7 &  5.0 &  1.55  &   $\pm$0.52      & -481.1   & ${+90.0}\atop{-188.3}$      &  22.91 & $\pm$5.56 \\
H95326(CCDMJ19236)& 0.0780&$\pm$0.0579 &  0.1017 & 0.0749 & -0.0216 & 0.0417 &  35.6 &  0.4 &  2.24  &  $\pm$26.40${^a}$& -341.3   & ${+123.5}\atop{-299.0}$     &  36.16 & ${+22.22}\atop{-0.62}$\\
H27887(GJ2046)    & 0.0780&$\pm$0.0005 & -0.0518 & 0.0006 & -0.0604 & 0.0006 &  30.4 &  0.2 &  2.02  &   $\pm$0.04      & -402.1   & $\pm$3.9                    &  30.79 & $\pm$0.22 \\
H8709(GJ3121)     & 0.0630&$\pm$0.0038 &  0.0012 & 0.0043 &  0.1241 & 0.0024 &  64.0 &  3.0 &  2.29  &   $\pm$0.33      & -237.4   & $\pm$19.5                   &  64.68 & $\pm$3.34 \\
H12351(GJ1049)    & 0.0610&$\pm$0.0013 & -0.0190 & 0.0013 &  0.0303 & 0.0013 &  26.2 & 10.0 &  1.74  &   $\pm$3.16${^a}$& -604.5   & ${+188.0}\atop{-640.4}$     &  26.37 & $\pm$10.96 \\
GJ54.2B(HD7438)   & 0.0574&$\pm$0.0009 &  0.1320 & 0.0022 &  0.2830 & 0.0021 &  16.0 &  5.0 & 14.80  &   $\pm$1.38      & -296.5   & ${+78.7}\atop{-24.5}$       &  30.36 & ${+3.53}\atop{-2.23}$ \\
H32362(GJ242)     & 0.0570&$\pm$0.0008 & -0.1152 & 0.0007 & -0.1909 & 0.0005 & 211.1 &  0.9 &  1.54  &   $\pm$0.05      &  -80.6   & $\pm$1.2                    & 211.91 & $\pm$1.01 \\
GJ401B(LHS290)    & 0.0569&$\pm$0.0252 & -1.8660 & 0.0249 & -0.6610 & 0.0184 &  40.2 &  6.5 & 17.08  & ${+251.76}\atop{-5.07}$ &  -24.0   & $\pm$10.4                   & 169.76 & ${+2511.15}\atop{-43.50}$ \\
H27288(GJ217.1)   & 0.0460&$\pm$0.0007 & -0.0148 & 0.0006 & -0.0012 & 0.0005 &  25.6 &  5.0 &  1.31  &   $\pm$0.38      & -827.1   & ${+142.6}\atop{-287.0}$     &  25.65 & $\pm$5.61 \\
H13772(HD18455)   & 0.0445&$\pm$0.0026 &  0.0302 & 0.0024 & -0.0369 & 0.0018 &  50.4 &  0.2 &  2.25  &   $\pm$0.30      & -431.4   & $\pm$26.6                   &  50.69 & $\pm$0.23 \\
H26373(HD37572)   & 0.0419&$\pm$0.0017 &  0.0253 & 0.0023 &  0.0000 & 0.0022 &  32.4 &  0.2 &  2.08  &   $\pm$0.25      & -714.4   & $\pm$31.1                   &  32.55 & $\pm$0.23 \\
T100111(HD351880) & 0.0400&$\pm$0.0327 & -0.0100 & 0.0340 &  0.0074 & 0.0270 &  26.1 &  0.3 &  1.40  & $\pm$70.48${^a}$& -933.4   & ${+370.5}\atop{-1163.6}$    &  26.15 & ${+20.85}\atop{-0.34}$ \\
H13769(GJ120.1C)  & 0.0389&$\pm$0.0015 &  0.0154 & 0.0015 & -0.0325 & 0.0012 &  49.6 &  0.5 &  2.26  &   $\pm$0.21      & -502.6   & $\pm$21.2                   &  49.83 & $\pm$0.56 \\
H93506(HD176687)  & 0.0370&$\pm$0.0014 & -0.0141 & 0.0026 &  0.0037 & 0.0020 &  22.0 &  5.0 &  2.25  &   $\pm$0.91      &-1192.9   & ${+231.3}\atop{-545.0}$     &  22.08 & $\pm$5.60 \\
H30067(HD43947)   & 0.0360&$\pm$0.0009 & -0.0175 & 0.0010 & -0.0143 & 0.0007 &  40.5 &  2.0 &  2.04  &   $\pm$0.19      & -667.1   & $\pm$41.7                   &  40.60 & $\pm$2.24 \\
S14576(ALGOL)     & 0.0350&$\pm$0.0009 &  0.0024 & 0.0008 & -0.0014 & 0.0009 &   4.0 &  0.9 &  2.54  &   $\pm$1.86${^a}$&-6893.3   & ${+1315.1}\atop{-2911.5}$   &   4.05 & $\pm$1.00 \\
H30344(HD44821)   & 0.0340&$\pm$0.0008 & -0.0030 & 0.0005 &  0.0041 & 0.0006 &  14.4 &  0.2 &  1.44  &   $\pm$0.17      &-1990.0   & $\pm$59.0                   &  14.44 & $\pm$0.23 \\
H54806(HD97578)   & 0.0310&$\pm$0.0141 & -0.0117 & 0.0095 &  0.0018 & 0.0099 &  23.5 &  1.0 &  2.48  &  $\pm$15.52${^a}$&-1333.5   & ${+371.4}\atop{-1355.2}$    &  23.59 & ${+1.64}\atop{-1.10}$ \\
H31626(HD260564)  & 0.0290&$\pm$0.0021 &  0.0095 & 0.0018 & -0.0332 & 0.0013 &  82.7 &  5.0 &  2.35  &   $\pm$0.44      & -405.8   & ${+35.4}\atop{-50.0}$       &  82.89 & $\pm$5.61 \\
H26624(HD37594)   & 0.0240&$\pm$0.0007 & -0.0039 & 0.0006 &  0.0021 & 0.0004 &  22.4 &  1.3 &  1.67  &   $\pm$0.27      &-1815.0   & $\pm$134.9                  &  22.43 & $\pm$1.46 \\
T31821(HD47787)   & 0.0210&$\pm$0.0021 & -0.0343 & 0.0017 &  0.0274 & 0.0025 &  18.3 &  0.6 & 22.66  &   $\pm$4.65      &-1966.5   & $\pm$128.1                  &  20.84 & $\pm$0.82 \\
H99483(HIP99483)  & 0.0130&$\pm$0.0067 &  0.0013 & 0.0214 &  0.0003 & 0.0251 &  25.0 &  0.2 &  1.95  & $\pm$129.44${^a}$&-3001.3   & $\pm$1154.8                 &  25.07 & ${+13.47}\atop{-0.18}$ \\
H40317(HD68814)   & 0.0120&$\pm$0.0016 & -0.0007 & 0.0014 &  0.0020 & 0.0012 &  34.2 &  0.2 &  2.02  &   $\pm$2.15${^a}$&-2378.2   & ${+275.3}\atop{-428.3}$     &  34.27 & $\pm$0.27 \\
H101573(HIP101573)& 0.0050&$\pm$0.0023 & -0.0002 & 0.0024 &  0.0004 & 0.0022 &  43.7 &  0.5 &  2.55  &  $\pm$21.96${^a}$&-4443.3   & $\pm$1418.0                 &  44.15 & ${+0.91}\atop{-0.59}$ \\

\hline
\end{tabular}
}\\ 
{\footnotetext{}{$^a$ The uncertainty value marked here corresponds
to the radius of a cylinder where the locus of the closest approach is
likely to lie.}}
\label{table.past}
\end{table*}

\begin{table*}
\caption{Astronomical data for the 33 stellar subsample run toward the future.}
\scalebox{0.9}[1.3]{
\begin{tabular}{@{}lr@{}lr@{$\pm$}lr@{$\pm$}lr@{$\pm$}lr@{}lr@{}lr@{}l@{}}

\hline
Star Name        & \multicolumn{2}{c}{Parallax}         & \multicolumn{2}{c}{PM($\alpha$)} & \multicolumn{2}{c}{PM($\delta$)} & \multicolumn{2}{c}{V$_r$}         & \multicolumn{2}{c}{Miss Distance} & \multicolumn{2}{c}{Time}         & \multicolumn{2}{c}{Vel$_{App}$} \\

& \multicolumn{2}{c}{(arcsec)}         & \multicolumn{2}{c}{(arcsec/yr)} & \multicolumn{2}{c}{(arcsec/yr)} & \multicolumn{2}{c}{(km/s)}         & \multicolumn{2}{c}{(pc)} & \multicolumn{2}{c}{(10$^3$ yr)}         & \multicolumn{2}{c}{(km/s)} \\
\hline

H70890(PROXIMA)   & 0.7723 & $\pm$0.0024 & -3.7756 & 0.0015 &  0.7682 & 0.0018 &  -21.7 & 0.5 & 0.95 & $\pm$0.01       &   26.7 &  $\pm$ 0.1              &  32.10 &  $\pm$0.39 \\
H71683(AlphCenA)  & 0.7421 & $\pm$0.0014 & -3.6200 & 0.0015 &  0.7100 & 0.0012 &  -20.7 & 0.9 & 1.01 & $\pm$0.02       &   27.7 &  ${+0.1}\atop{-0.2}$    &  31.37 &  $\pm$0.67 \\
H71681(AlphCenB)  & 0.7421 & $\pm$0.0014 & -3.6004 & 0.0261 &  0.9521 & 0.0198 &  -24.6 & 0.9 & 0.94 & $\pm$0.02       &   27.7 &  $\pm$0.2               &  34.22 &  $\pm$0.74 \\
H87937(BARNARD)   & 0.5490 & $\pm$0.0016 & -0.7978 & 0.0016 & 10.3269 & 0.0013 & -106.8 & 0.2 & 1.17 & $\pm$0.01       &    9.8 &  $\pm$0.1               & 139.30 &  $\pm$0.25 \\
H54035(GJ411)     & 0.3924 & $\pm$0.0009 & -0.5802 & 0.0008 & -4.7671 & 0.0008 &  -85.0 & 0.9 & 1.44 & $\pm$0.01       &   20.0 &  $\pm$0.1               & 102.91 &  $\pm$0.84 \\
S32349(SIRIUS)    & 0.3792 & $\pm$0.0016 & -0.5460 & 0.0013 & -1.2231 & 0.0012 &   -9.4 & 0.9 & 2.30 & $\pm$0.06       &   65.8 &  ${+3.0}\atop{-4.5}$    &  19.20 &  $\pm$0.50 \\
H92403(GJ729)     & 0.3365 & $\pm$0.0018 &  0.6376 & 0.0022 & -0.1925 & 0.0015 &   -4.0 & 2.0 & 2.73 & $\pm$0.22       &  111.8 &  ${+24.6}\atop{-57.9}$  &  10.20 &  ${+1.12}\atop{-0.55}$ \\
GJ905(LHS549)     & 0.3160 & $\pm$0.0020 &  0.0850 & 0.0050 & -1.6150 & 0.0050 &  -81.0 & 5.0 & 0.91 & $\pm$0.06       &   35.1 &  $\pm$2.1               &  84.56 &  $\pm$5.39 \\
H57548(GJ447)     & 0.2996 & $\pm$0.0022 &  0.6056 & 0.0021 & -1.2192 & 0.0019 &  -31.0 & 0.2 & 1.91 & $\pm$0.03       &   71.0 &  $\pm$0.3               &  37.75 &  $\pm$0.21 \\
GJ866A(LHS68)     & 0.2895 & $\pm$0.0050 &  2.3640 & 0.0050 &  2.2360 & 0.0050 &  -60.0 & 2.0 & 2.29 & $\pm$0.08       &   31.5 &  ${+0.1}\atop{-0.2}$    &  80.24 &  $\pm$1.82 \\
H104214(GJ820A)   & 0.2871 & $\pm$0.0015 &  4.1551 & 0.0010 &  3.2589 & 0.0012 &  -64.3 & 0.9 & 2.80 & $\pm$0.03       &   18.7 &  $\pm$0.1               & 108.34 &  $\pm$0.73 \\
H110893(GJ860A)   & 0.2495 & $\pm$0.0030 & -0.8702 & 0.0030 & -0.4711 & 0.0030 &  -24.0 & 5.0 & 2.47 & $\pm$0.36       &  101.2 &  ${+2.2}\atop{-7.2}$    &  30.49 &  $\pm$4.37 \\
H85605(CCDMJ)     & 0.2027 & $\pm$0.0395 &  0.0973 & 0.0267 &  0.3489 & 0.0413 &  -21.1 & 0.2 & 1.84 & ${+1.20}\atop{-0.47}$ &  197.0 &  ${+36.5}\atop{-25.4}$  &  22.74 &  ${+1.18}\atop{-0.52}$ \\
H86214(GJ682)     & 0.1983 & $\pm$0.0024 & -0.7101 & 0.0028 & -0.9380 & 0.0021 &  -60.0 &10.0 & 2.14 & $\pm$0.36       &   67.4 &  $\pm$7.8               &  66.27 &  $\pm$10.12 \\
H97649(GJ768)     & 0.1944 & $\pm$0.0009 &  0.5368 & 0.0007 &  0.3855 & 0.0007 &  -26.1 & 0.9 & 2.70 & $\pm$0.08       &  139.5 &  $\pm$2.4               &  30.68 &  $\pm$0.86 \\
H57544Ã‚Â (GJ445) & 0.1855 & $\pm$0.0014 &  0.7432 & 0.0016 &  0.4804 & 0.0012 & -119.0 & 5.0 & 1.01 & $\pm$0.05       &   43.0 &  $\pm$1.9               & 121.13 &  $\pm$5.53 \\
H86990(GJ693)     & 0.1721 & $\pm$0.0022 & -1.1199 & 0.0021 & -1.3525 & 0.0015 & -115.0 &21.0 & 2.25 & $\pm$0.44       &   42.0 &  ${+7.0}\atop{-4.9}$    & 124.76 &  $\pm$21.61 \\
H99461(GJ783A)    & 0.1652 & $\pm$0.0009 &  0.4569 & 0.0009 & -1.5749 & 0.0006 & -129.8 & 0.9 & 2.06 & $\pm$0.03       &   40.3 &  $\pm$0.3               & 138.07 &  $\pm$0.96 \\
H86961(GJ2130A)   & 0.1618 & $\pm$0.0113 & -0.0498 & 0.0929 & -0.3198 & 0.0473 &  -28.9 & 0.9 & 1.93 & $\pm$0.41       &  188.8 &  $\pm$12.4              &  30.42 &  $\pm$1.09 \\
H86963(GJ2130B)   & 0.1618 & $\pm$0.0113 & -0.0776 & 0.0136 & -0.2701 & 0.0078 &  -27.4 & 0.9 & 1.78 & $\pm$0.28       &  202.3 &  $\pm$13.7              &  28.61 &  $\pm$0.97 \\
H83945(GJ3991)    & 0.1378 & $\pm$0.0090 &  0.3339 & 0.0081 & -0.2780 & 0.0103 &  -45.0 &10.0 & 2.29 & $\pm$0.71       &  142.0 &  ${+37.8}\atop{-22.1}$  &  47.42 &  $\pm$10.58 \\
H93449(RCrA)      & 0.1218 & $\pm$0.0682 & -0.0344 & 0.0975 &  0.0506 & 0.0520 &  -36.0 & 5.0 & 0.54 & $\pm$27.21${^a}$&  222.0 &  ${+462.7}\atop{-76.4}$ &  36.08 &  ${+14.33}\atop{-5.55}$ \\
H77257Ã‚Â (GJ598) & 0.0851 & $\pm$0.0008 & -0.2255 & 0.0008 & -0.0685 & 0.0007 &  -66.4 & 0.9 & 2.28 & $\pm$0.06       &  166.5 &  $\pm$2.8               &  67.69 &  $\pm$1.00 \\
H116727(GJ903)    & 0.0725 & $\pm$0.0005 & -0.0489 & 0.0005 &  0.1272 & 0.0004 &  -42.4 & 0.9 & 2.84 & $\pm$0.08       &  305.0 &  $\pm$6.9               &  43.33 &  $\pm$0.99 \\
H6379(GJ56.5)     & 0.0595 & $\pm$0.0006 & -0.0341 & 0.0005 & -0.0345 & 0.0006 &  -22.7 & 2.0 & 2.82 & $\pm$0.29       &  703.4 &  ${+79.7}\atop{-55.7}$  &  23.03 &  $\pm$2.22 \\
H89825(GJ710)     & 0.0518 & $\pm$0.0014 &  0.0017 & 0.0014 &  0.0021 & 0.0011 &  -13.9 & 2.0 & 0.34 & $\pm$0.28${^a}$ & 1357.8 &  ${+312.1}\atop{-172.7}$&  13.90 &  $\pm$2.25 \\
H113421(HD217107) & 0.0507 & $\pm$0.0008 & -0.0061 & 0.0008 & -0.0160 & 0.0006 &  -14.0 & 0.6 & 2.22 & $\pm$0.16       & 1355.5 &  $\pm$67.0              &  14.16 &  $\pm$0.67 \\
H38228(HD63433)   & 0.0458 & $\pm$0.0009 & -0.0093 & 0.0010 & -0.0118 & 0.0007 &  -16.5 & 0.2 & 2.08 & $\pm$0.16       & 1281.6 &  $\pm$31.0              &  16.58 &  $\pm$0.22 \\
H105766(GJ4194)   & 0.0389 & $\pm$0.0006 &  0.0412 & 0.0006 &  0.0396 & 0.0006 &  -76.9 & 0.2 & 2.32 & $\pm$0.07       &  324.2 &  $\pm$5.3               &  77.22 &  $\pm$0.23 \\
H20359(GJ168)     & 0.0326 & $\pm$0.0020 & -0.0348 & 0.0018 &  0.0111 & 0.0015 &  -78.5 & 5.0 & 2.07 & $\pm$0.34       &  380.3 &  $\pm$37.0              &  78.68 &  $\pm$5.61 \\
H21386(HD26367)   & 0.0273 & $\pm$0.0015 &  0.0135 & 0.0013 &  0.0090 & 0.0017 &  -50.7 & 2.0 & 2.01 & $\pm$0.29       &  704.1 &  $\pm$51.3              &  50.80 &  $\pm$2.25 \\
H85661(HD158576)  & 0.0115 & $\pm$0.0008 &  0.0010 & 0.0009 &  0.0000 & 0.0004 &  -46.0 & 1.7 & 0.97 & $\pm$0.89${^a}$ & 1847.7 &  $\pm$157.1             &  46.02 &  $\pm$1.89 \\
H94512(HD1779939) & 0.0085 & $\pm$0.0009 & -0.0001 & 0.0007 & -0.0005 & 0.0005 &  -30.1 & 2.0 & 1.49 & $\pm$1.95${^a}$ & 3821.3 &  $\pm$480.8             &  30.10 &  $\pm$2.20 \\
\hline
\end{tabular}
}\\ 
{\footnotetext{}{$^a$ The uncertainty value marked here corresponds
to the radius of a cylinder where the locus of the closest approach is
likely to lie.}}
\label{table.future}
\end{table*}

\subsubsection{The Most Interesting Candidates}\label{interesting_candidates_sample}

From this sample, if we ignored the uncertaities, we could conlcude
that none of these stars would approach to the Sun closer than
$\sim70,000$ AU (more than 200 times beyond than necessary to start
affecting the outskirts of our planetary system). We have considered
however, observational uncertainties in this work. The stars that may
produce important (even dangerous) approaches to the Sun, are those
whose tangential velocity were potentially zero (simultaneously in
$\alpha$ and $\delta$). We will then check on those stars with a
tangential velocity (or approach distance to the Sun), is consistent
with zero at 3$\sigma$. This includes both stars with small tangential
velocity and those with uncertain tangential velocities. This makes a
total number of 12, 8 to the past and 4 to the future, from the whole
sample of 67 objects.

\begin{figure}
\includegraphics[width=90mm]{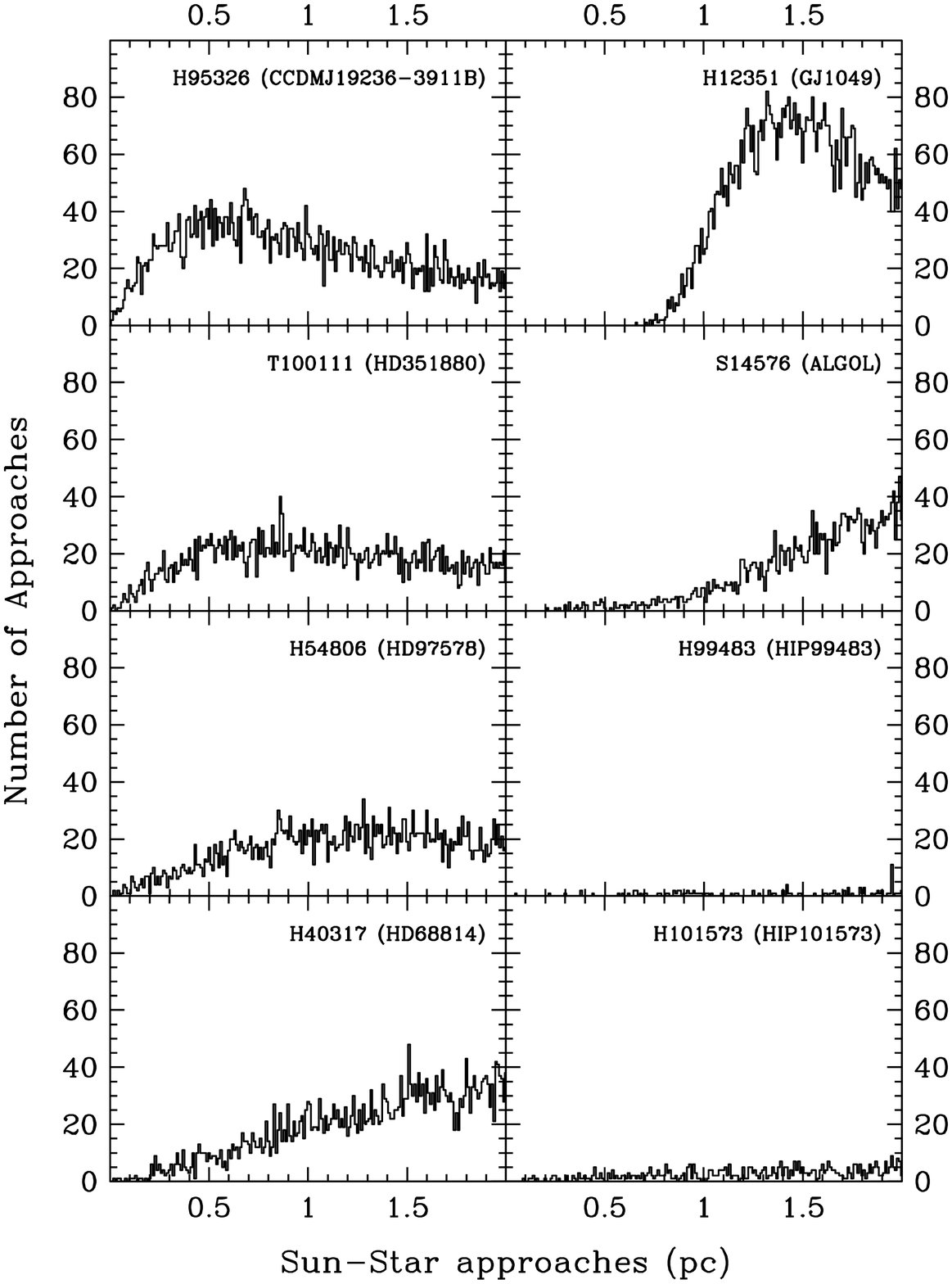}
\vspace{0.1cm}
\caption {Histograms computed on eight stars from our sample to the
  past with large enough observational error bars that may produce
  approach distances between the Sun and the star below zero. All
  plots show the Sun-star approaches in pc from a set of 10,000 normal
  deviation random initial conditions within observational error
  bars.}
\label{fig.histpast}
\end{figure}

\begin{figure}
\includegraphics[width=90mm]{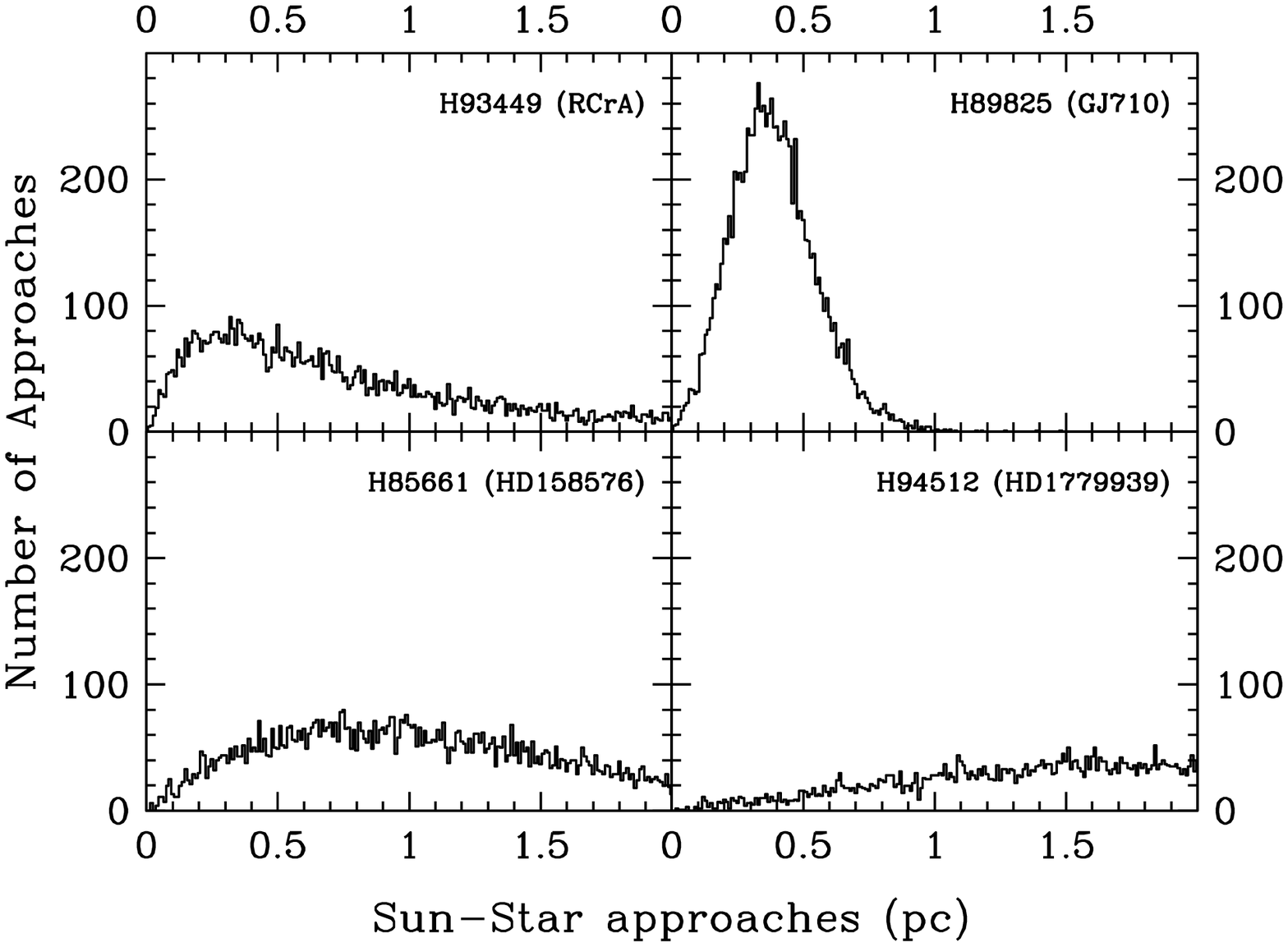}
\vspace{0.1cm}
\caption {Same as Figure \ref{fig.histpast} but for four stars to the future.}
\label{fig.histfut}
\end{figure}

For these 12 stars (labeled with an asterisc in Tables
\ref{table.past} and \ref{table.future}), we have separately produced
in Figures \ref{fig.histpast} and \ref{fig.histfut}, histograms, eight
to the past and four to the future. In each histogram we computed
10,000 random realizations with a normal distribution for the initial
conditions within all the observational uncertainties, with the
notable exception of the paralax, where the value was restricted to
positive values. In these diagrams we analize the most likely approach
distance sun-star toward the past and future respectively. All
diagrams were done with the same bin size of 0.01 pc and the same axes
(for all stars in the corresponding diagram) for better see
differences between the stellar approaches to the past or to the
future.

The number of approaches closer than 0.01 pc is of 3 to the past and 8
to the future, this is equivalent to a probability of 0.03\% to the
past and 0.08\% to the future of an approach to less than 0.01 pc. A
more interesting distance is of course one that could cause a
noticeable effect on the planetary disc, that we know now is
approximately of 300 AU, as we will show in Section
\ref{solarsystem_solarneigbourhood}. The probability of a closer than
300 AU approach is then of $\sim0.0017\%$ from the 8-stars to the past
and of $\sim0.0006\%$ from the 4-stars to the future. The probabilities
with these stars to pass close enough to affect the planetary disc of
the Solar system are extremely low, even for the star that will
approach closest (GJ 710).

\subsubsection{Comparison between the Straight Line Approximation, 
a Local, and Global Potentials}\label{globalpotential_versus_others}

Following Garc\'ia-S\'anchez et al. (2001), we compare stellar orbits
with our global potential to those using the straight line
approximation and a local potential approximation from their
paper. For distances of the stars larger than 50 pc the local and
straight line approximations give important differences with respect to
a global model, even with times as short as 1 Myr.

Garc\'ia-S\'anchez et al. (2001) concluded that within a time interval
of $\pm$ 10 Myr from the present time, the predicted encounters are
fairly well determined for most of the candidate stars. They are not
altered significantly by the use of alternative Galactic potential
models or by varying the plausible values of the Galactic
parameters. The most interesting result is the future passage of
Gliese 710 through the outer Oort cloud. This result is in good
agreement with the predictions using other Galactic potential models,
the prediction of this stellar passage is not model dependent owing to
its proximity to the Sun. They integrated the trajectories using three
different models of the Galactic potential: a local potential model, a
global potential model and a perturbative potential model. The
agreement between their models was generally good.

We are interested in knowing the approximate distance to the stars,
where the straight line and local approximations differ significantly
from the global model. To do so, we use the data in Table 2 of
Garc\'ia-S\'anchez et al. (1999) for the closest encounters, using a
simple rectilinear motion of the stars, and also the data for the
closest encounters using a local Galactic potential.

\begin{table*}
\centering
\caption{Relative errors between different approximations to the
  Galactic potential. Column 1 is the name of the object; Columns 2, 3
  and 4 show the close-approach distance to the Sun in pc with the global
  (from this work), local and straight line approximations
  (Garc\'ia-S\'anchez et al. 2001,1999); Columns 5,6 and 7 present the
  relative errors (G=Global, Lo=Local, Li=Linear) and the last column
  is the current distance to the star in pc.}
\begin{tabular}{@{}llclccccclllllll@{}}
\hline
\hline
Star Name  &Global        &Local            &Linear       &(G-Lo)/G&(G-Li)/G&(Lo-Li)/Lo&  Dist             \\
               &       & Missing Distance         &             &               &               &               &                               \\
               &       & (pc)        &             &               &               (\%)      &         &                            (pc)    \\
\hline
GJ 710	        &	0.337	&	0.336	&	0.343	&	0.3	&	1.8	&	2.1	&	19.30		\\
HD 158576	&	0.938	&	0.846	&	0.753	&	9.8	&	19.7	&	11.0	&	86.81		\\
PROXIMA	        &	0.954	&	0.954	&	0.954	&	0.0	&	0.0	&	0.0	&	1.29	        \\
ALPHA CENTAURI A	&	0.973	&	0.973	&	0.973	&	0.0	&	0.0	&	0.0	&	1.35		\\
ALPHA CENTAURI B	&	0.975	&	0.975	&	0.975	&	0.0	&	0.0	&	0.0	&	1.35		\\
AC + 79 3888	&	1.007	&	1.007	&	1.007	&	0.0	&	0.0	&	0.0	&	5.39		\\
GJ 620.1B	&	4.259	&	1.139	&	1.139	&	73.3	&	73.3	&	0.0	&	12.81	        \\
BARNARD STAR	&	1.144	&	1.143	&	1.143	&	0.1	&	0.1	&	0.0	&	1.82		\\
HD 351880	&	1.434	&	1.439	&	1.445	&	0.3	&	0.8	&	0.4	&	25.27		\\
LALANDE 21185	&	1.440	&	1.440	&	1.440	&	0.0	&	0.0	&	0.0	&	2.55		\\
SAO 75395	&	2.469	&	1.448	&	2.688	&	41.4	&	8.9	&	85.6	&	118.34		\\
HD 179939	&	1.444	&	1.451	&	1.025	&	0.5	&	29.0	&	29.4	&	117.10		\\
GJ 208	        &	1.600	&	1.600	&	1.599	&	0.0	&	0.1	&	0.1	&	11.38		\\
HD 37594	&	1.637	&	1.610	&	1.598	&	1.6	&	2.4	&	0.7	&	41.39		\\
GJ 217.1	&	1.645	&	1.637	&	1.629	&	0.5	&	1.0	&	0.5	&	21.52		\\
HIP 99483	&	1.797	&	1.653	&	1.379	&	8.0	&	23.3	&	16.6	&	74.13		\\
HD 35317	&	1.775	&	1.755	&	1.735	&	1.1	&	2.3	&	1.1	&	58.04		\\
GJ 2130 B	&	1.782	&	1.782	&	1.782	&	0.0	&	0.0	&	0.0	&	6.18		\\
HD 19995	&	1.254	&	1.811	&	2.653	&	44.4	&	111.6	&	46.5	&	68.54		\\
HIP 101573	&	2.072	&	1.821	&	1.898	&	12.1	&	8.4	&	4.2	&	187.62		\\
CCDM 17296 + 2439 B	&	1.837	&	1.837	&	1.837	&	0.0	&	0.0	&	0.0	&	4.93		\\
GJ 358	        &	1.875	&	1.875	&	1.875	&	0.0	&	0.0	&	0.0	&	9.49		\\
ROSS 154	&	1.881	&	1.881	&	1.881	&	0.0	&	0.0	&	0.0	&	2.97		\\
HD 68814	&	1.950	&	1.909	&	1.990	&	2.1	&	2.1	&	4.2	&	82.10		\\
ROSS 128	&	1.911	&	1.911	&	1.911	&	0.0	&	0.0	&	0.0	&	3.34		\\
GJ 2130 A	&	1.929	&	1.929	&	1.929	&	0.0	&	0.0	&	0.0	&	6.18		\\
GJ 860 A	&	1.949	&	1.949	&	1.949	&	0.0	&	0.0	&	0.0	&	4.01		\\
HD 33487	&	2.001	&	1.977	&	1.954	&	1.2	&	2.3	&	1.2	&	41.63		\\
HD 43947	&	2.015	&	2.016	&	2.016	&	0.0	&	0.0	&	0.0	&	27.53		\\
HD 26367	&	2.019	&	2.028	&	2.038	&	0.4	&	0.9	&	0.5	&	36.66		\\
GJ 271 A	&	2.044	&	2.038	&	2.029	&	0.3	&	0.7	&	0.4	&	18.03		\\
GJ 168	        &	2.074	&	2.074	&	2.075	&	0.0	&	0.0	&	0.0	&	30.69		\\
GJ 144	        &	2.135	&	2.135	&	2.135	&	0.0	&	0.0	&	0.0	&	3.22		\\
HD 63433	&	2.150	&	2.138	&	2.121	&	0.6	&	1.3	&	0.8	&	21.82		\\
GJ 682	        &	2.140	&	2.140	&	2.140	&	0.0	&	0.0	&	0.0	&	5.04		\\
GJ 120.1	&	2.243	&	2.245	&	2.246	&	0.1	&	0.1	&	0.0	&	22.48		\\
GJ 693	        &	2.253	&	2.253	&	2.253	&	0.0	&	0.0	&	0.0	&	5.81		\\
CCDM 19236 - 3911 B	&	2.262	&	2.261	&	2.260	&	0.0	&	0.1	&	0.0	&	12.87		\\
HD 122676	&	2.264	&	2.263	&	2.262	&	0.0	&	0.1	&	0.0	&	26.12		\\
GJ 598	        &	2.267	&	2.267	&	2.267	&	0.0	&	0.0	&	0.0	&	11.75		\\
GJ 120.1 C	&	2.267	&	2.269	&	2.269	&	0.1	&	0.1	&	0.0	&	25.73		\\
WD 0148+467	&	2.286	&	2.286	&	2.286	&	0.0	&	0.0	&	0.0	&	15.85		\\
SIRIUS	        &	2.299	&	2.299	&	2.299	&	0.0	&	0.0	&	0.0	&	2.64		\\
HD 37574	&	2.290	&	2.305	&	2.233	&	0.7	&	2.5	&	3.1	&	62.00		\\
HD 217107	&	2.300	&	2.313	&	2.323	&	0.6	&	1.0	&	0.4	&	19.72		\\
HD 176687	&	2.299	&	2.314	&	2.333	&	0.7	&	1.5	&	0.8	&	27.31		\\
BD -02 3986	&	1.804	&	2.316	&	3.102	&	28.4	&	72.0	&	33.9	&	58.48		\\
HD 67852	&	2.932	&	2.341	&	1.229	&	20.2	&	58.1	&	47.5	&	118.34		\\
HD 260564	&	2.341	&	2.341	&	2.340	&	0.0	&	0.0	&	0.0	&	34.42		\\
ALGOL	        &	2.481	&	2.381	&	2.666	&	4.0	&	7.5	&	12.0	&	28.46		\\
GJ 54.1	        &	2.429	&	2.429	&	2.429	&	0.0	&	0.0	&	0.0	&	3.72		\\
LP 816 - 60	&	2.482	&	2.482	&	2.483	&	0.0	&	0.0	&	0.0	&	5.49		\\
HD 50867	&	2.540	&	2.587	&	2.732	&	1.9	&	7.6	&	5.6	&	51.47		\\
GJ 16	        &	2.596	&	2.609	&	2.623	&	0.5	&	1.0	&	0.5	&	16.22		\\
HD 34790	&	2.477	&	2.647	&	2.862	&	6.9	&	15.5	&	8.1	&	85.32		\\
IRAS 17249+0416	&	2.670	&	2.664	&	2.658	&	0.2	&	0.4	&	0.2	&	50.08		\\
HD 152912	&	2.860	&	2.700	&	2.466	&	5.6	&	13.8	&	8.7	&	139.28		\\
GJ 768	        &	2.702	&	2.702	&	2.702	&	0.0	&	0.0	&	0.0	&	5.14		\\
GJ 903	        &	2.791	&	2.791	&	2.792	&	0.0	&	0.0	&	0.0	&	13.79		\\
HD 172748	&	2.801	&	2.806	&	2.823	&	0.2	&	0.8	&	0.6	&	57.34		\\
GJ 56.5	        &	2.825	&	2.823	&	2.823	&	0.1	&	0.1	&	0.0	&	16.82		\\
GJ 908	        &	2.886	&	2.886	&	2.885	&	0.0	&	0.0	&	0.0	&	5.97		\\
HD 221420	&	2.917	&	2.907	&	2.886	&	0.3	&	1.1	&	0.7	&	31.76		\\
\hline
\end{tabular}
\label{table.relative_errors}
\end{table*}

\begin{table*}
\centering
\caption{countinued...}
\begin{tabular}{@{}lcccccccccccccccccccccccccccccccccc@{}}
\hline
\hline
Star Name &Global        &Local            &Linear       &(G-Lo)/G&(G-Li)/G&(Lo-Li)/Lo&  Dist             \\
               &       & Missing Distance         &             &               &               &               &                               \\
               &       & (pc)        &             &               &               (\%)      &         &                            (pc)    \\
\hline
HD 142500	&	3.247	&	2.917	&	2.458	&	10.2	&	24.3	&	15.7	&	73.86		\\
ROSS 614	&	2.929	&	2.929	&	2.929	&	0.0	&	0.0	&	0.0	&	4.12		\\
GJ 1095	        &	2.969	&	2.969	&	2.968	&	0.0	&	0.0	&	0.0	&	16.86		\\
ROSS 882	&	3.053	&	3.052	&	3.052	&	0.0	&	0.0	&	0.0	&	5.93		\\
HD 148317	&	3.268	&	3.132	&	2.903	&	4.2	&	11.2	&	7.3	&	79.87        	\\
HD 150689	&	3.145	&	3.145	&	3.146	&	0.0	&	0.0	&	0.0	&	14.37		\\
GJ 169	        &	3.189	&	3.189	&	3.188	&	0.0	&	0.0	&	0.0	&	11.47		\\
GJ 279	        &	3.196	&	3.196	&	3.197	&	0.0	&	0.0	&	0.0	&	25.70		\\
GJ 628	        &	3.208	&	3.208	&	3.208	&	0.0	&	0.0	&	0.0	&	4.26		\\
GJ 687	        &	3.213	&	3.213	&	3.213	&	0.0	&	0.0	&	0.0	&	4.53		\\
GJ 231	        &	3.249	&	3.249	&	3.249	&	0.0	&	0.0	&	0.0	&	10.15		\\
HD 38382	&	3.271	&	3.271	&	3.273	&	0.0	&	0.1	&	0.1	&	25.54		\\
GJ 71	        &	3.271	&	3.271	&	3.271	&	0.0	&	0.0	&	0.0	&	3.65		\\
GJ 1005	        &	3.289	&	3.289	&	3.289	&	0.0	&	0.0	&	0.0	&	5.21		\\
VAN MAANEN STAR	&	3.327	&	3.327	&	3.327	&	0.0	&	0.0	&	0.0	&	4.41		\\
HD 28676	&	3.681	&	3.380	&	2.966	&	8.2	&	19.4	&	12.2	&	39.14		\\
GJ 722	        &	3.384	&	3.384	&	3.384	&	0.0	&	0.0	&	0.0	&	12.98		\\
HD 233081	&	3.412	&	3.396	&	3.355	&	0.5	&	1.7	&	1.2	&	51.02		\\
GJ 280 A	&	3.438	&	3.438	&	3.438	&	0.0	&	0.0	&	0.0	&	3.50		\\
GJ 15A	        &	3.467	&	3.469	&	3.469	&	0.1	&	0.1	&	0.0	&	3.57		\\
GJ 678	        &	3.503	&	3.503	&	3.503	&	0.0	&	0.0	&	0.0	&	16.45		\\
GJ 725 B	&	3.515	&	3.515	&	3.515	&	0.0	&	0.0	&	0.0	&	3.52		\\
HD 199881	&	4.355	&	3.527	&	3.106	&	19.0	&	28.7	&	11.9	&	80.45		\\
HD 67523	&	3.563	&	3.563	&	3.564	&	0.0	&	0.0	&	0.0	&	19.23		\\
GJ 725A	        &	3.568	&	3.568	&	3.568	&	0.0	&	0.0	&	0.0	&	3.57		\\
GJ 66	        &	3.569	&	3.570	&	3.570	&	0.0	&	0.0	&	0.0	&	8.15		\\
HD 168769	&	3.540	&	3.594	&	3.662	&	1.5	&	3.4	&	1.9	&	50.33		\\
LUYTEN STAR	&	3.666	&	3.666	&	3.666	&	0.0	&	0.0	&	0.0	&	3.80		\\
GJ 825	        &	3.696	&	3.696	&	3.696	&	0.0	&	0.0	&	0.0	&	3.95		\\
GJ 784	        &	3.727	&	3.727	&	3.727	&	0.0	&	0.0	&	0.0	&	6.20		\\
GJ 775	        &	3.756	&	3.756	&	3.756	&	0.0	&	0.0	&	0.0	&	13.11		\\
GJ 96	        &	3.756	&	3.756	&	3.756	&	0.0	&	0.0	&	0.0	&	11.91		\\
GJ 251	        &	3.813	&	3.814	&	3.814	&	0.0	&	0.0	&	0.0	&	5.52		\\
BD + 37 4901C	&	3.940	&	3.820	&	3.622	&	3.0	&	8.1	&	5.2	&	33.80		\\
GJ 380	        &	3.856	&	3.856	&	3.856	&	0.0	&	0.0	&	0.0	&	4.87		\\
GJ 252	        &	3.869	&	3.867	&	3.862	&	0.1	&	0.2	&	0.1	&	17.27		\\
HD 122064	&	3.868	&	3.868	&	3.868	&	0.0	&	0.0	&	0.0	&	10.10		\\
HD 168956	&	4.134	&	3.940	&	3.762	&	4.7	&	9.0	&	4.5	&	73.69		\\
HD 53253	&	3.527	&	3.998	&	4.381	&	13.4	&	24.2	&	9.6	&	125.16		\\
HD 146214	&	4.394	&	4.034	&	4.007	&	8.2	&	8.8	&	0.7	&	93.63		\\
GJ 791.1A	&	4.021	&	4.053	&	4.103	&	0.8	&	2.0	&	1.2	&	30.27		\\
GJ 268	        &	4.066	&	4.066	&	4.066	&	0.0	&	0.0	&	0.0	&	6.36		\\
HD 192869	&	3.979	&	4.080	&	4.072	&	2.5	&	2.3	&	0.2	&	111.98		\\
HD 33959C	&	4.092	&	4.093	&	4.097	&	0.0	&	0.1	&	0.1	&	25.14		\\
GJ 337.1	&	4.117	&	4.121	&	4.121	&	0.1	&	0.1	&	0.0	&	19.56		\\
GJ 674	        &	4.134	&	4.134	&	4.134	&	0.0	&	0.0	&	0.0	&	4.54		\\
GJ 620.1A	&	4.153	&	4.155	&	4.158	&	0.0	&	0.1	&	0.1	&	12.87		\\
GJ 103	        &	4.180	&	4.180	&	4.180	&	0.0	&	0.0	&	0.0	&	11.51		\\
GJ 851	        &	4.205	&	4.203	&	4.203	&	0.0	&	0.0	&	0.0	&	11.44		\\
HD 170296	&	4.256	&	4.278	&	4.280	&	0.5	&	0.6	&	0.0	&	89.37		\\
GJ 222	        &	4.380	&	4.380	&	4.380	&	0.0	&	0.0	&	0.0	&	8.66		\\
GJ 752A	        &	4.420	&	4.420	&	4.420	&	0.0	&	0.0	&	0.0	&	5.87		\\
HD 218200	&	4.856	&	4.429	&	4.062	&	8.8	&	16.4	&	8.3	&	74.91		\\
HD 207164	&	2.720	&	4.449	&	7.785	&	63.6	&	186.2	&	75.0	&	75.87		\\
GJ 688	        &	4.461	&	4.461	&	4.460	&	0.0	&	0.0	&	0.0	&	10.71         	\\
HD 58954	&	4.467	&	4.483	&	4.343	&	0.4	&	2.8	&	3.1	&	85.98		\\
GJ 716	        &	4.484	&	4.484	&	4.484	&	0.0	&	0.0	&	0.0	&	13.21		\\
HD 32450	&	4.488	&	4.489	&	4.490	&	0.0	&	0.0	&	0.0	&	8.52		\\
HD 162102	&	4.430	&	4.518	&	4.629	&	2.0	&	4.5	&	2.5	&	64.68		\\
GJ 68 	        &	4.572	&	4.573	&	4.573	&	0.0	&	0.0	&	0.0	&	7.47		\\
HD 71974	&	4.639	&	4.612	&	4.564	&	0.6	&	1.6	&	1.0	&	28.71		\\
HD 163547	&	4.056	&	4.648	&	5.170	&	14.6	&	27.5	&	11.2	&	149.93		\\
\hline
\end{tabular}
\label{table.relative_errors}
\end{table*}

\begin{table*}
\centering
\caption{continued...}
\begin{tabular}{@{}lcccccccccccccccccccccccccccccccccc@{}}
\hline
\hline
Star Name &Global        &Local            &Linear       &(G-Lo)/G&(G-Li)/G&(Lo-Li)/Lo&  Dist             \\
               &       & Missing Distance         &             &               &               &               &                               \\
               &       & (pc)        &             &               &               (\%)      &         &                            (pc)    \\
\hline
ROSS 780	&	4.690	&	4.690	&	4.690	&	0.0	&	0.0	&	0.0	&	4.70		\\
GJ 702	        &	4.698	&	4.698	&	4.698	&	0.0	&	0.0	&	0.0	&	5.09		\\
GJ 178	        &	4.701	&	4.701	&	4.701	&	0.0	&	0.0	&	0.0	&	8.03		\\
GJ 701	        &	4.720	&	4.720	&	4.720	&	0.0	&	0.0	&	0.0	&	7.80		\\
HD 72617	&	4.760	&	4.762	&	4.748	&	0.0	&	0.3	&	0.3	&	58.07		\\
GJ 832	        &	4.828	&	4.828	&	4.828	&	0.0	&	0.0	&	0.0	&	4.94		\\
GJ 638	        &	4.834	&	4.834	&	4.834	&	0.0	&	0.0	&	0.0	&	9.77		\\
GJ 713	        &	4.838	&	4.838	&	4.838	&	0.0	&	0.0	&	0.0	&	8.06		\\
HD 239927	&	4.862	&	4.871	&	4.890	&	0.2	&	0.6	&	0.4	&	57.44		\\
GJ 625	        &	4.896	&	4.896	&	4.896	&	0.0	&	0.0	&	0.0	&	6.58		\\
HD 67228	&	4.924	&	4.924	&	4.925	&	0.0	&	0.0	&	0.0	&	23.33		\\
GJ 735	        &	4.926	&	4.927	&	4.927	&	0.0	&	0.0	&	0.0	&	11.59		\\
HD 75935	&	4.857	&	4.938	&	5.063	&	1.7	&	4.2	&	2.5	&	40.55		\\
GJ 410	        &	4.973	&	4.976	&	4.980	&	0.1	&	0.1	&	0.1	&	11.66		\\
HD 39655	&	4.694	&	4.979	&	5.197	&	6.1	&	10.7	&	4.4	&	102.77		\\
GJ 644	        &	4.982	&	4.982	&	4.982	&	0.0	&	0.0	&	0.0	&	5.74		\\
HD 120702	&	4.820	&	4.993	&	5.184	&	3.6	&	7.6	&	3.8	&	96.25		\\
\hline
\end{tabular}
\label{table.relative_errors}
\end{table*}

\begin{figure}
\includegraphics[width=90mm]{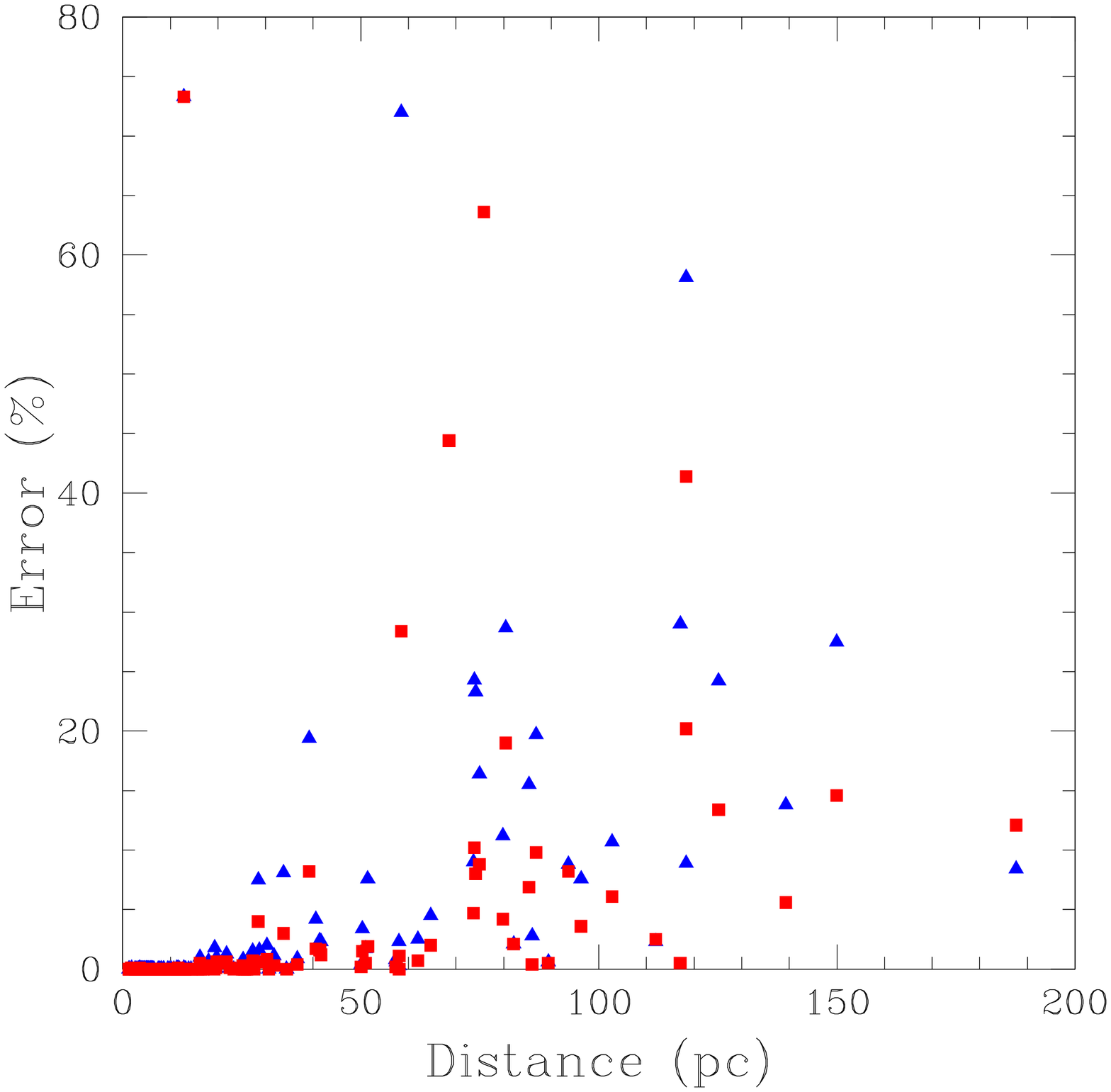}
\vspace{0.1cm}
\caption {Relative error between the global and the straight line
approximations to the Galactic potential (blue triangles) and between
the global and local approximations (red squares).}
\label{fig.error}
\end{figure}

In Table \ref{table.relative_errors}: we show 142 stars close to the
Sun (within approximately 190 pc). From this sample, we obtain 74
($\sim 52\%$) stars for which the difference among the approximations
to the potential employed (global, local or linear) is negligible. In
Figure \ref{fig.error}, we show the relative errors of the closest
approaches (missing distance errors), between the global {\it vs.}
straight line approximations (blue triangles), and between the global
{\it vs.}  local approximations (red squares), as a function of
distance from the Sun. From this figure, we can appreciate that the
three approximations give the same results up to distances of $\sim$20
pc. Even for times as short as 10 Myr, stars beyond $\sim$20 pc have
significant differences in their trajectories for the different
approximations, showing where the Galactic global potential becomes
important. A local potential clearly does better than the straight
line approximation. For completeness, we compare the results of our
global model including arms and bar, with the simple global model of
Garc\'ia-S\'anchez et al. (2001), that includes spiral arms. In this
case we find 73$\%$ of our trajectories differ by less than 3\%, and
for the rest (the most distant stars in general) the error is
6-26$\%$. Differences among stellar orbits due to the use of different
models for the Galactic field or the simple stright-line
approximation, become more important the farer the star is, making
even more uncertain approaches to the Sun in its path for the Milky
Way disc. Deeper and precise observations as the ones are coming in
the near future, will improve dramaticaly our knowledge on this field,
and will make more important to use better models of the Galaxy.

\subsubsection{The Solar System Under the Influence of Stellar Encounters at the Solar Neighborhood}\label{solarsystem_solarneigbourhood}
The overarching goal of our work in this set of papers will be to
address quantitatively the effect of stellar encounters in different
Galactic environments. In this section, we start with the nascent
Solar System and the current stellar environment.  For this purpose, we
model a simple planetary system (or disk) with 1000 particles
distributed from 1 to 100 AU. We study this system for a total
integration time of 10,000 yr, which is much longer that the typical
encounter timescale.

We chose the parameters of Gliese 710 (0.6 M$_{\odot}$, approximation
velocity 13.9 km s$^{-1}$) and an estimated closest approach less than
0.3 pc, 1.4 Myr in the future, to calculate its effects in the Solar
System. The effect of this specific star will be interesting for the
Oort cloud, but we find that Gliese 710's effect is negligible for a
100 AU planetary system.

\begin{figure}
\includegraphics[width=84mm]{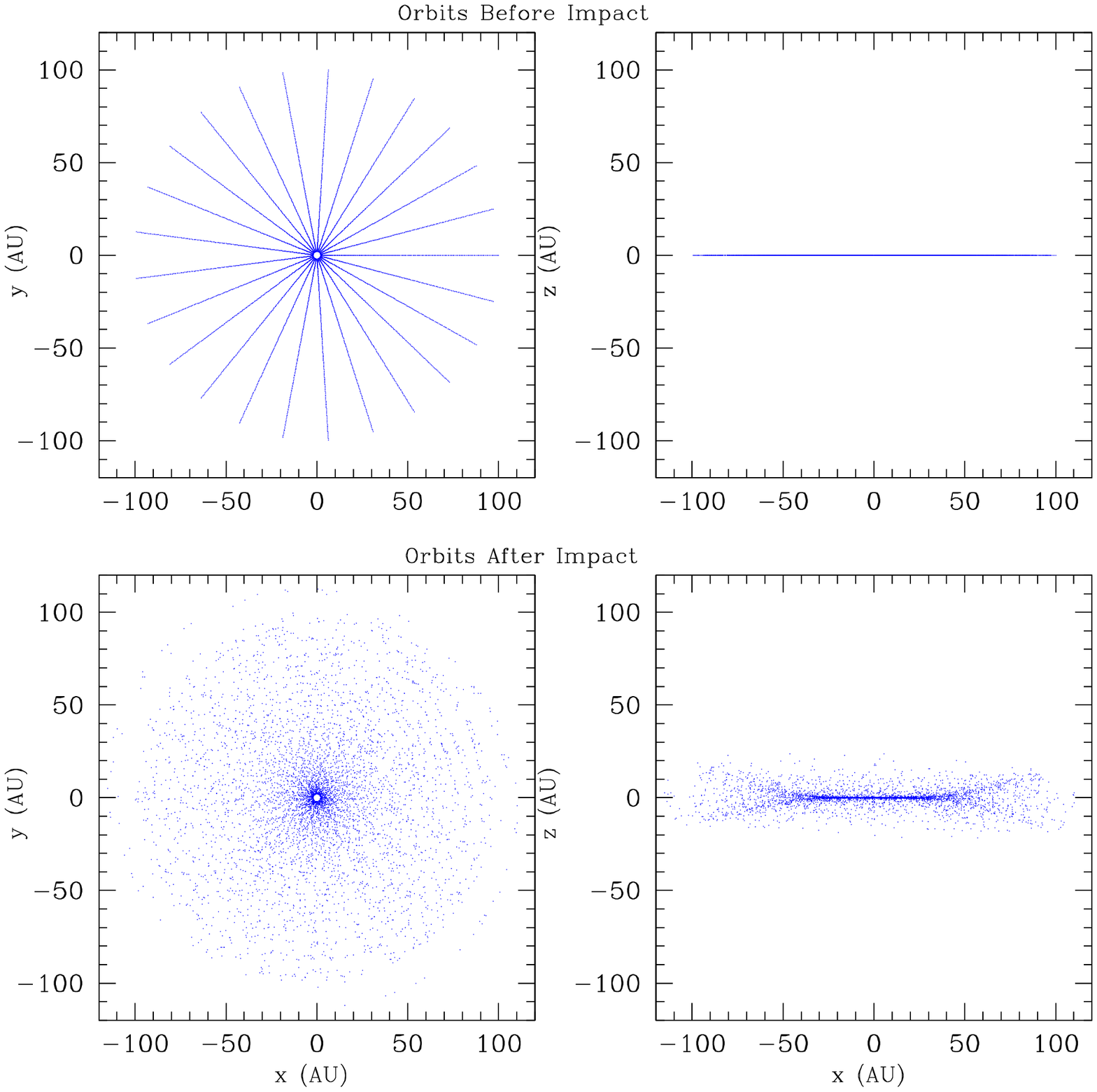}
\vspace{0.1cm}
\caption {Effect of a stellar flyby mass of 1 $M_{\odot}$, closest
  approach of 300 AU and velocity of 13.9 km s$^{-1}$ on a disc of
  particles. Left frames: orbits of the particles in x-y cuts, right
  frames: cut in the x-z plane. Upper frames: particles before the
  encounter, lower frames: particles after the encounter}
\label{fig.orbitasperturbador}
\end{figure}

\begin{figure}
\includegraphics[width=84mm]{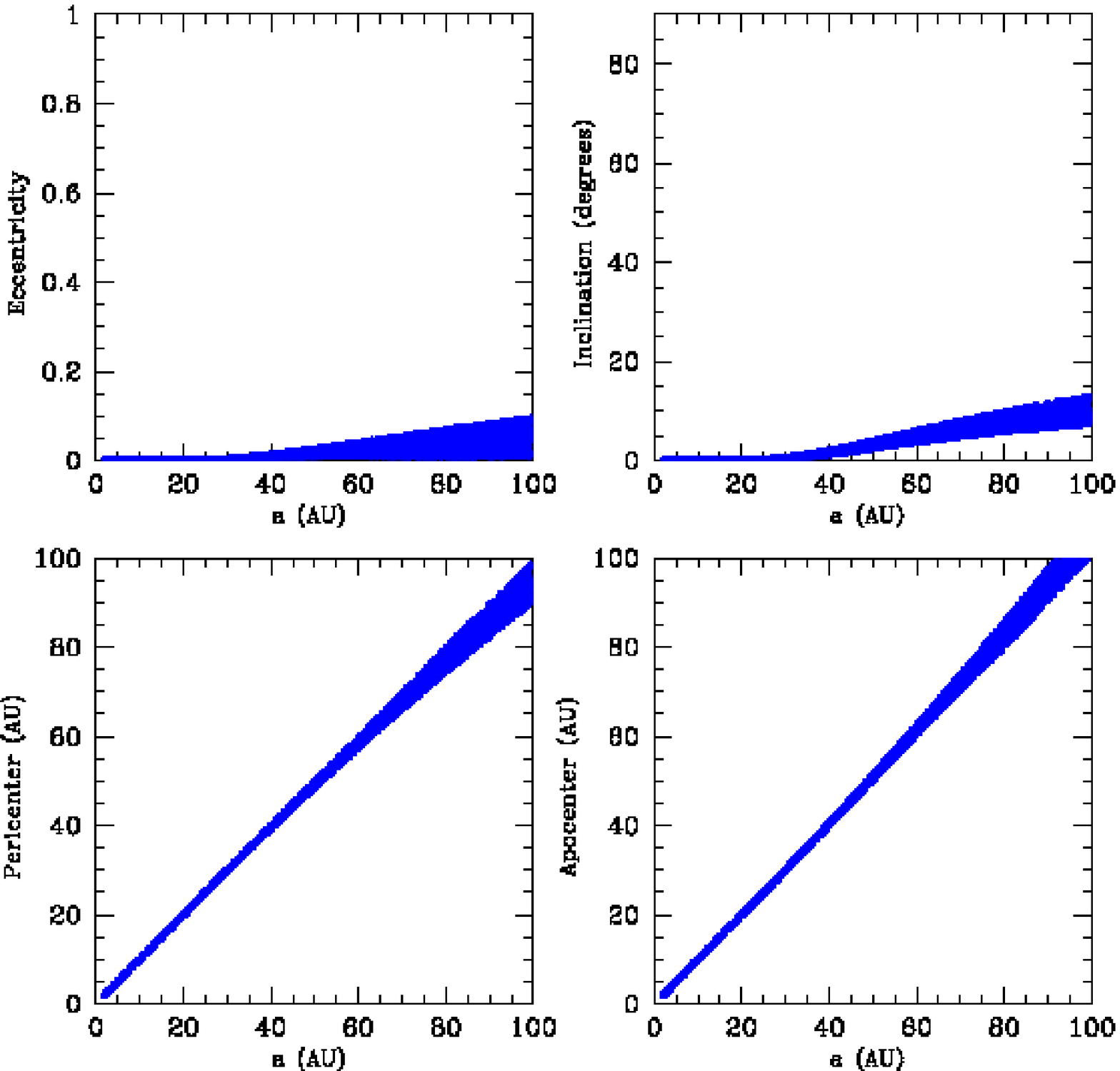}
\vspace{0.1cm}
\caption {Resultant orbital characteristics on the disc of Figure
  \ref{fig.orbitasperturbador} after the encounter. Upper left frame:
  eccentricity, upper right frame: inclination, lower left frame:
  pericenter distance, and lower right frame: apocenter distance. All
  versus the semimajor axis}
\label{fig.paramperturbador}
\end{figure}

We ran a grid of simulations for a flyby star mass of 1 M$_{\odot}$
and closest approaches between 100 AU and 1000 AU, in steps of 50
AU. Disk disruption starts to be significant for approaches within
about 200 AU. We also find a good agreement to the analytical result
of Hall, Clarke \& Pringle (1996) where they find that a disk is
affected to $\sim 1/3$ the closest approach distance. With these
parameters and the velocity of GJ 710 (13.9 km s$^{-1}$), we calculate
the gravitational effect on a planetary disc and we present it in
Figure \ref{fig.orbitasperturbador}. The figure shows four panels,
representing the orbits of particles in x-y (left frames) and x-z
(right frames) planes, before (upper frames) and after (lower
frames). Figure \ref{fig.paramperturbador} shows the resultant orbital
characteristics of the disc after the encounter, eccentricity of
particles (upper left frame), inclination (upper right frame),
pericenter distance (lower left frame) and apocenter distance (lower
right frame), all plotted versus the semimajor axis. The effect on the
disc is slight but clear starting at 40 AU, where particles reach an
eccentricity up to 0.1.

\subsubsection{Effect of Gliese 710 on the Oort Cloud}
The solar system's disk and Kuiper belt are surrounded by the Oort
cloud, which contains $1 \times 10^{11}$ to $5 \times 10 ^{12}$
cometary nuclei with a total mass $\sim$ 1 to 50 $M_{\oplus}$ (Stern
2003). Comets in the Oort cloud evolve dynamically under the influence
of external perturbers such as random stellar passages. Close or
penetrating passages of stars through the Oort cloud can deflect
comets toward the inner planetary region (Hills 1981; Weissman 1996;
Kirsh \etal 2009; Brasser \etal 2006, 2007, 2008).

One important issue in the scenario of the flyby star is the effect it
would have on the loosely bound outer Oort cloud. Small gravitational
perturbations could have severe effects on the cloud due to the direct
effect on the Sun. A simple calculation using the impulse
approximation predicts that during a stellar encounter the velocity of
the Sun would change by $\delta$$v$ $\approx$ $ \frac {2 \ G \ {M_*}}
{{q_*} \ {v_{\infty}}} $. With the values of Gliese 710, for the
stellar mass, $M_{*}$ = 0.6 M$_{\odot}$, the closest distance of the
encounter, $q_{*}$ = 0.34 pc ($\sim$70,000 AU) and the velocity at
infinity, $v_{\infty}$=13.4 km s$^{-1}$. This would result in $5.8
\times 10^{-4}$ km s$^{-1}$ change in the solar velocity, a negligible
value compared with the typical Oort cloud speed of 0.2 km s$^{-1}$ or
with its escape velocity at the boundary ($\sim 0.1$ km s$^{-1}$), so
Gliese 710 has no ability to strip the Oort cloud from the
Sun. However, it has the potential to send a comet flux toward the
inner Solar system. This will cause the appearance of a new comet per
year and the net increase of the risk of astronomic impact will not be
detectable (Garc\'ia-S\'anchez et al. 2001).

\section{Second Galactic Region: The Birth Cloud of the Sun}
\label{sun_birth_cloud_simulations}

In star formation regions such as Sun's putative birth location inside
a dense cluster, stellar densities are high enough to produce stellar
encounters within 300 AU before the dissolution of the stellar cluster
(Laughlin \& Adams 1998; Adams 2010). Also, in this environment,
encounters are stronger owing to lower typical velocity dispersions
between 1 and 3 km s$^{-1}$. Let us consider, for example, a planetary
disc around a given star on a crowded stellar environment. Let us
assume solar mass stars for generality. The star-disc system will
experience encounters with other systems with an interaction rate that
can be written as $\gamma = \langle n\sigma v \rangle = \langle n
\rangle \langle \sigma \rangle \langle v \rangle$, where $n$ is the
number density of stars in the cluster, $\sigma$ is the cross section
for interaction, and $v$ is the relative velocity (velocity dispersion
in this case). For a rough estimate of the interaction rate among
stars, we use the typical values for open clusters. In their work,
Laughlin \& Adams (1998), calculate the interaction rate for the
Trapezium cluster, using a central density of $n_0 \approx 5\times
10^4$ pc$^{-3}$ and a velocity dispersion of a few km s$^{-1}$, and an
interaction cross section of 100$^2$ AU$^2$, obtaining a rate of
interaction of one encounter every 40 million years. Considering open
clusters live for at least 100 million years, it is expected these
interactions are significant in those environments. In the case of the
Sun's birth cloud several observables, such as the inclinations of
Uranus and Neptune (which are sensitive to stellar interactions) place
densities at similar amounts of typical open clusters, this is $\sim
3\times 10^4$ M$_\odot$ pc$^3$ (Gaidos 1995). Furthermore, the orbit
of (90377) Sedna supports the idea that the Solar system was born in a
stellar cluster with a non-negligible density (Brasser \etal 2006;
Kenyon \& Bromley 2004; Morbidelli \& Levison 2004).

In this section, we present experiments to reproduce some of the Kuiper
belt bodies orbital characteristics, such as eccentricities and
inclinations. We have chosen four close-approach distances (200, 150,
100 and 50 AU), coupled with three initial approach velocities (1, 2
and 3 km s$^{-1}$) for a stellar flyby interacting with a 100 AU disc
of particles. The mass of the perturbing star is 1 M$_{\odot}$, and
angles involved in the geometry of stellar encounter have the general
values, $\theta$=$45^{\circ}$ (angle between the plane of the stellar
trajectory and the plane of the disc) and $\alpha$=$45^{\circ}$ (angle
between the axis of the stellar trajectory plane and the axis of the
disc).

\begin{figure*}
\includegraphics[width=125mm]{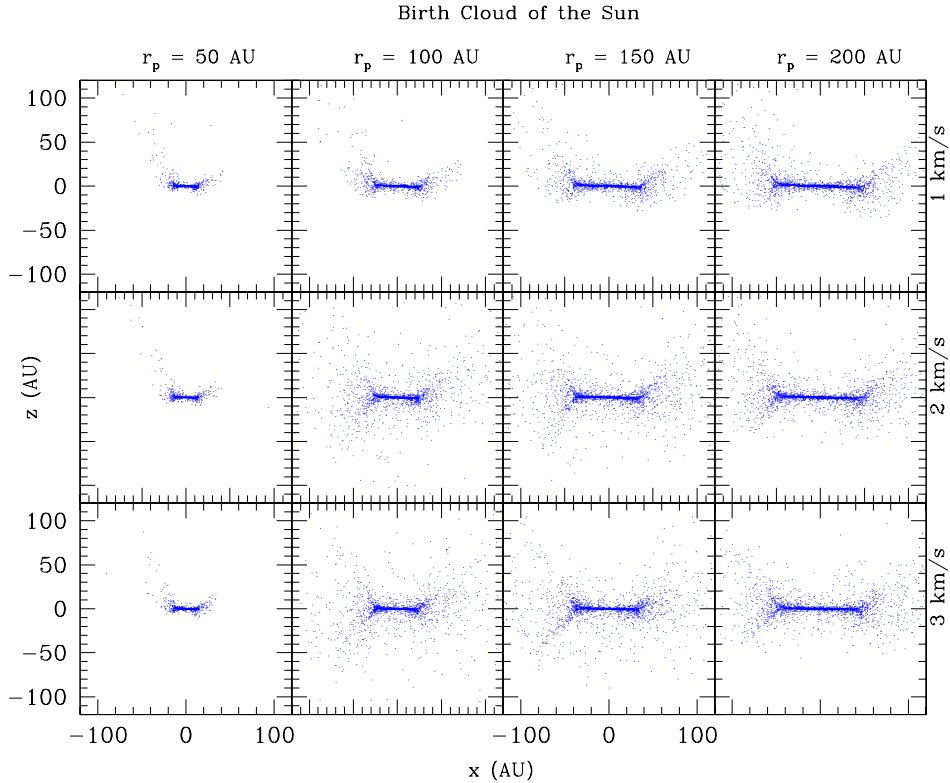}
\vspace{0.1cm}
\caption {Final disc structure after a stellar encounter simulating a
  Galactic environment of star formation. Rows represent values of
  velocity dispersion within 3 km s$^{-1}$ and columns represent
  different impact parameters within 200 AU}
\label{fig.orbits_experiments}
\end{figure*}

In our initial conditions, test particles have $e=i=0$. In Figure
\ref{fig.orbits_experiments} we show our results of perturbations
produced in the disk of particles by a flyby. The columns represent
different values of the initial approach velocity (1, 2 and 3 km
s$^{-1}$) and rows represent different values of closest approach
distances (50,100,150 and 200 AU). Each panel shows the positions of
the test particles in the x-z plane after the encounter with the flyby
star.

\begin{figure*}
\includegraphics[width=125mm]{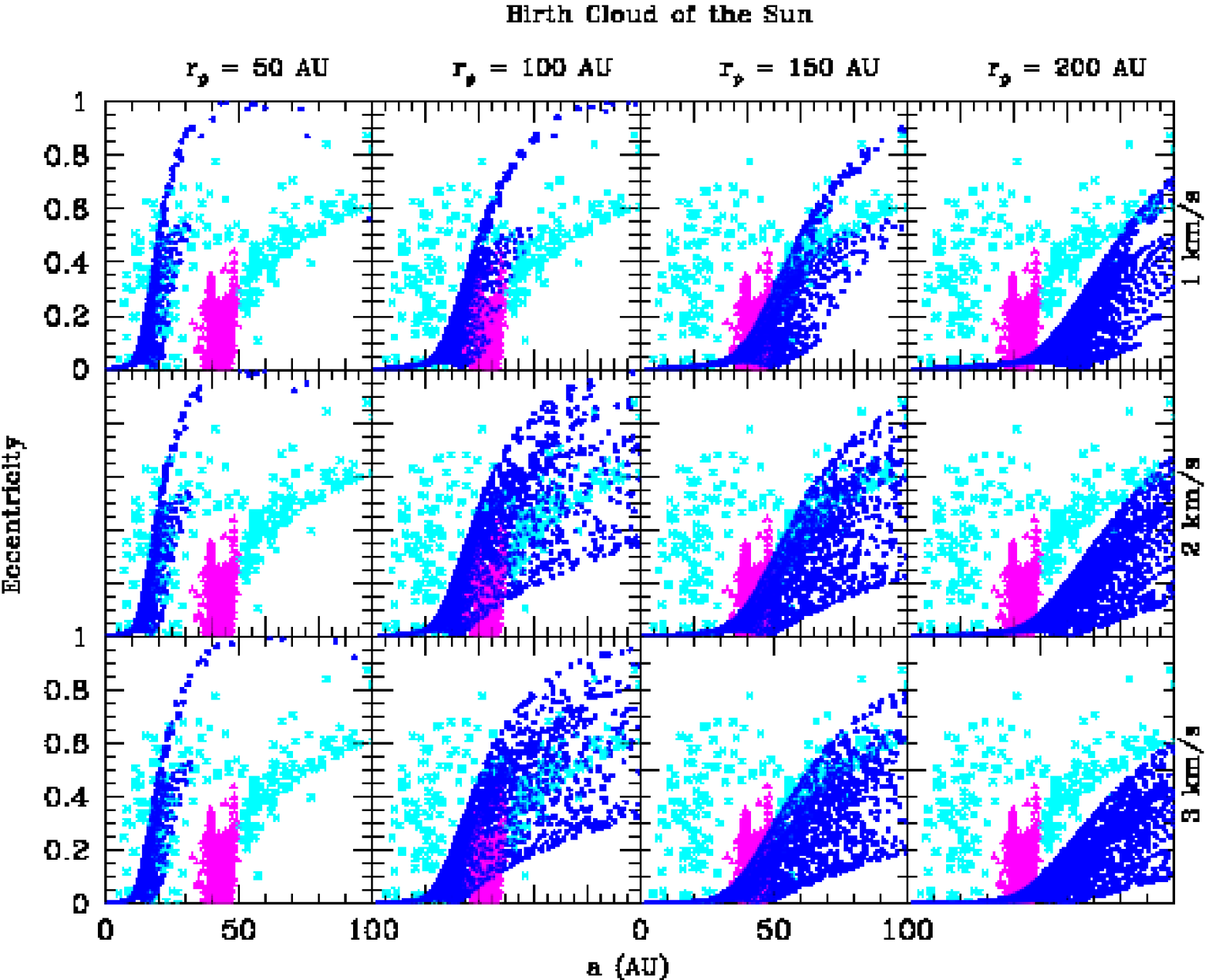}
\vspace{0.1cm}
\caption {Resultant eccentricities. Rows indicate different values of
velocity dispersion and columns represent different distances of
closest approach. Our best fit happens for a stellar encounter with
minimum distance of 100-150 AU and a velocity of $\sim$1 km
s$^{-1}$. Resonant objects and classic Kuiper belt objects are
included (pink triangles), and also scattered objects and Centaurs at
radii less that 30 AU (cyan crosses)}
\label{fig.eccentricities_experiments}
\end{figure*}

\begin{figure*}
\includegraphics[width=125mm]{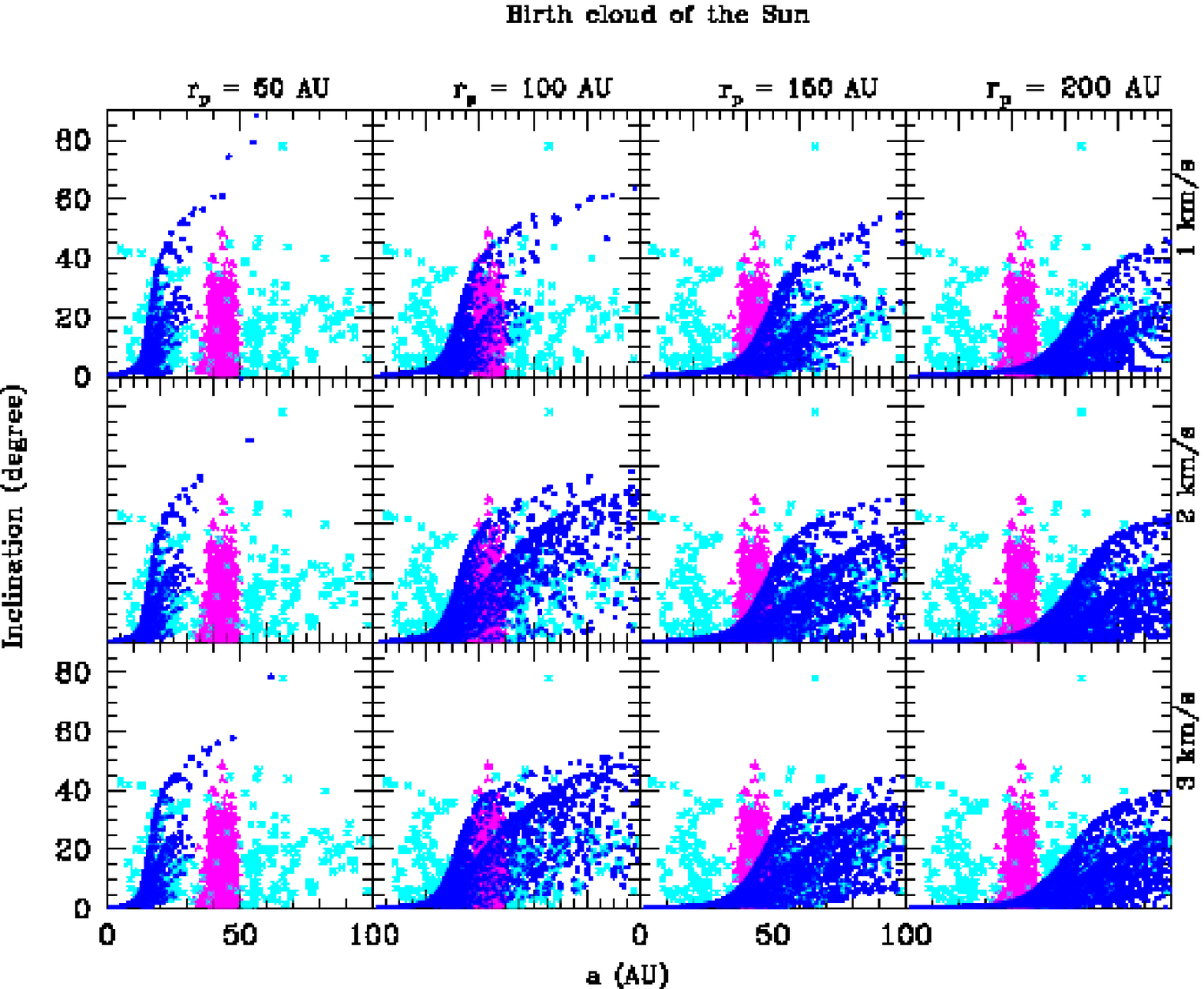}
\vspace{0.1cm}
\caption {Same as Figure \ref{fig.eccentricities_experiments} but for
inclinations}
\label{fig.inclinations_experiments}
\end{figure*}

In Figures \ref{fig.eccentricities_experiments} and
\ref{fig.inclinations_experiments} we show the resultant
eccentricities and inclinations, plotted versus semi-major
axis. Columns represent different distances of closest approach,
within 200 AU, and rows indicate different values of velocity
dispersion, within 3 km s$^{-1}$. As a reference, in all plots, we
include the known Kuiper belt objects, including the resonant and
classic objects (pink triangles), and the scattered objects, including
also Centaurs at radii less than 30 AU (green crosses).

Our best fit happens at a distance of maximum approach between 100 and
150 AU and a velocity of $\sim$1 km s$^{-1}$, with an unaltered inner
planetary system (as far as 30 AU) where the most of particles between
30 and 50 AU have up to $e = 0.2$ and $i < 20^{\circ}$.  In this
simulation we also obtain some dispersed objects with high
eccentricities and low semimajor axis as many of the objects seen in
the planetary system. From 42 to approximately 48 AU there are objects
with eccentricities up to 0.4. We are able to match these objects
using encounters with a velocity at infinity of 1 km s$^{-1}$ and a
close-approach distance of 100 AU. This results in eccentricities from
0-0.1, on semimajor axis interval 0-40 AU, and eccentricities from 0
to 1 on semimajor axis in the interval 40-65 AU and from 65 until 100
AU the most of particles are dynamically evaporated, this means that
objects with low eccentricity and large semimajor axis (larger than 50
or 60 AU) are scarce in these experiments as it is the case in the
Kuiper belt.

\section{Conclusions}\label{conclusions}

This paper is the first step in an extensive study of the
gravitational effect of a flyby star on a 100 AU disc, representing
debris discs (and/or planetary systems) given the particular
conditions of several environments of the Galaxy. In this work, we
analyzed the effect of the Solar neighborhood and the birth cloud of
the Sun. In the current Solar neighborhood, we find that Gliese 710
will be the closest star to the Sun 1.36 Myr in the future with a
minimum distance of 0.34 pc ($\sim$70,000 AU). This stellar encounter
will lead to direct interactions between the star and the Oort
cloud. This star has a mass of 0.6 M$_{\odot}$ currently located at
19.3 pc from the Sun. We calculated its velocity at the point of
minimum distance with the Sun as 14 km s$^{-1}$. The effect generated
by Gliese 710 on a 100 AU planetary system will be negligible. The
minimum distance for which the effect caused by Gliese 710 would start
to be noticeable is $\sim$300 AU, producing a slight heating of the
outer parts of the 100 AU planetary disc, with eccentricities up to
0.1 on semimajor axes between 60-100 AU and inclinations up to
10$^{\circ}$. With a simple impulse approximation we find that Gliese
710 will not even have an important effect on the global structure of
the Oort Cloud.

For stars of the Solar neighborhood, with 6D known information, we
constructed orbits in order to compare the use of a simple straight
line approximation, a local approximation to the Galactic potential,
an axisymmetric potential and a full Galactic potential (including
spiral arms and bar observationally motivated). Even for times as
short as 10 Myr, stars beyond $\sim$20 pc have significant differences
in their trajectories for the different approximations, showing where
the Galactic global potential becomes important. A local potential
clearly does better than the straight line approximation. Comparing
the global potential model (including spiral arms and bar), with the
simpler model of Garc\'ia-S\'anchez et al. (2001) (that includes
spiral arms), we find that 73$\%$ of our trajectories differ by less
than 3\%, and for the rest (the most distant stars in general) the
error is 6-26$\%$. Differences among stellar orbits due to the use of
different models for the Galactic potential field or the simple
straight line approximation, become more important the farer the star
is, making even more uncertain approaches to the Sun in its path for
the Milky Way disc. To compute precise stellar orbits of the Solar
neighborhood, taking advantage of the new generation of data produced
by large surveys, will requiere better models of the Milky Way Galaxy
instead of simpler approximations.

Finally, regarding the birth cloud of the Sun, we produced several
experiments to reproduce the orbital parameters of the Kuiper belt
objects. We know that at 42-48 AU there are objects with
eccentricities up to 0.4. We are able to approximate these objects
using encounters with a velocity at infinity of 1 km s$^{-1}$ and a
close-approach distance of 100 AU. This results in eccentricities from
0 - 0.1, on semimajor axis interval 0-40 AU, and eccentricities from 0
to 1 on semimajor axis in the interval 40-65 AU and from 65 until 100
AU the most of particles are dynamically evaporated.

\section*{Acknowledgments}
We are grateful to the anonymous referee for valuable comments which
contributed to improve the present paper. B.P. and J.J. thank Antonio
Peimbert, Justin Read, and Prasenjit Saha, for enlightening
discussions and Santiago Torres-Rodriguez for his help with the
stellar catalogues. J.J and B.P. thank projects UNAM through grants
PAPIIT IN110711-2 and IN-112210-2 and CONACYT through grant 60581.


\label{lastpage}


\begin{thebibliography}{}

\bibitem[\protect\citeauthoryear{Adams}{2010}]{A10} 
Adams, F.C. 2010, \araa, 48, 47

\bibitem[\protect\citeauthoryear{Antoja et al.}{2009}]{AV09} Antoja,
T.; Valenzuela, O.; Pichardo, B.; Moreno, E.; Figueras, F.;
Fern\'andez, D. 2009, \apj, 700, 78

\bibitem[Binney (2007)]{BT} 
Binney, J., Tremaine, S., 2007, {\it Galactic Dynamics}, Princeton
University Press

\bibitem{2010AstL...36..220B} Bobylev, V.~V.\ 2010, {\it Astronomy
Letters}, 36, 220

\bibitem[\protect\citeauthoryear{Bower}{2009}]{2009ApJ...701...1922B}
Bower, G.; Bolatto, A.; Ford, E.; and Kalas, P.;
2009, \apj, 701,1922

\bibitem[\protect\citeauthoryear{Brasser \etal}{2006}]{BDL06}
Brasser, R.; Duncan, M. J.; Levison, H. F. 2006, Icar, 184, 59

\bibitem[\protect\citeauthoryear{Brasser \etal}{2007}]{BDL07}
Brasser, R.; Duncan, M. J.; Levison, H. F. 2007, Icar, 191, 413

\bibitem[\protect\citeauthoryear{Brasser \etal}{2008}]{BDL08}
Brasser, R.; Duncan, M. J.; Levison, H. F. 2008, Icar, 196, 274

\bibitem[Carpenter(2000)]{2000AJ....120.3139C} 
Carpenter, J.~M.\ 2000, \aj, 120, 3139

\bibitem[Clark et al.(2005)]{2005prpl.conf.8171C} Clark, P.~C.,
Bonnell, I.~A., Zinnecker, H., \& Bate, M.~R.\ 2005, Protostars and
Planets V, 8171

\bibitem[Chakrabarty(2007)]{2007A&A...467..145C} Chakrabarty, D.\ 
2007, \aap, 467, 145

\bibitem[Dehnen(2000)]{2000AJ....119..800D} Dehnen, W.\ 2000, \aj, 119, 800 

\bibitem[\protect\citeauthoryear{FuenteMarcos}{1997}]{FM} 
de la Fuente Marcos C., de la Fuente Marcos R., 1997, A$\&$A, 326, L21

\bibitem[Dybczy{\'n}ski(2006)]{2006A&A...449.1233D} Dybczy{\'n}ski, P.~A.\ 2006, \aap, 449, 1233 

\bibitem[\protect\citeauthoryear{Gaidos}{1995}]{G95}
Gaidos, E. J. 1995, Icar, 114, 258

\bibitem[\protect\citeauthoryear{garciasanchez}{1997}]{GS97}
Garc\'\i a-S\'anchez, J., Preston, R., Jones, D., \etal 1997, in
Procedding of the ESA Symp. Hipparcos-venice 97, ed. B. Battrick (ESA
SP-402, Noordwijk), 617

\bibitem[\protect\citeauthoryear{garciasanchez}{1999}]{GS99}
Garc\'\i a-S\'anchez, J., Preston, R., Jones, D., \etal 1999, AJ, 117,
1042

\bibitem[Garc{\'{\i}}a-S{\'a}nchez et al.(2001)]{2001A&A...379..634G}
Garc{\'{\i}}a-S{\'a}nchez, J., Weissman, P.~R., Preston, R.~A., Jones,
D.~L., Lestrade, J.-F., Latham, D.~W., Stefanik, R.~P., \& Paredes,
J.~M.\ 2001, \aap, 379, 634

\bibitem[\protect\citeauthoryear{Gilmore}{1992}]{G} 
Gilmore, G., 1992, In {\it The Astronomy and Astrophysics
Encyclopedia}, p.643, ed. Maran S. P. Cambridge University Press

\bibitem[Goswami \& Vanhala(2000)]{2000prpl.conf..963G} Goswami,
J.~N., \& Vanhala, H.~A.~T.\ 2000, Protostars and Planets IV, 963

\bibitem[Hall et al.(1996)]{1996MNRAS.278..303H} Hall, S.~M., Clarke, C.~J., \& Pringle, J.~E.\ 1996, \mnras, 278, 303 

\bibitem[Hester et al.(2004)]{2004Sci...304.1116H} Hester, J.~J.,
Desch, S.~J., Healy, K.~R., \& Leshin, L.~A.\ 2004, Science, 304, 1116

\bibitem[\protect\citeauthoryear{Hills}{1981}]{H}
Hills, J., 1981, AJ, 86, 1730-1740

\bibitem[\protect\citeauthoryear{Hurleyshara}{2002}]{HS}
Hurley, J., Shara, M., 2002, ApJ, 565, 1251

\bibitem[Ida et al.(2000)]{2000ApJ...528..351I} 
Ida, S., Larwood, J., \& Burkert, A.\ 2000, \apj, 528, 351

\bibitem[\protect\citeauthoryear{Kenyon \& Bromley}{2004}]{KB04}
Kenyon, S.; Bromley, B., 2004, Natur, 432, 598

\bibitem[\protect\citeauthoryear{Kirsh \etal}{2009}]{KDL09}
Kirsh, D.; Duncan, M.; Brasser, R.; Levison, H.. 2009, Icar, 199, 197

\bibitem[Kobayashi \& Ida(2001)]{2001Icar..153..416K} 
Kobayashi, H., \& Ida, S.\ 2001, Icarus, 153, 416

\bibitem[\protect\citeauthoryear{Lada \& Lada}{2003}]{LL06}
Lada, C.J. \& Lada, E.A. 2003, ARA\&A, 41, 57

\bibitem[Laughlin \& Adams(1998)]{1998ApJ...508L.171L} 
Laughlin, G., \& Adams, F.~C.\ 1998, \apjl, 508, L171

\bibitem[\protect\citeauthoryear{Looney}{2006}]{L06}
Looney, L., Tobin, J., Fields, B., 2006, ApJ, 652, 1755-1762.

\bibitem[Martos et al.(2004)]{2004MNRAS.350L..47M} Martos, M., Hernandez, X., Y{\'a}{\~n}ez, M., Moreno, E., \& Pichardo, B.\ 2004, \mnras, 350, L47 

\bibitem[\protect\citeauthoryear{Mathews}{1994}]{M94}
Mathews, R., 1994, QJRAS, 35, 1(erratum 35, 243)

\bibitem[Meyer \& Clayton(2000)]{2000SSRv...92..133M} 
Meyer, B.~S., \& Clayton, D.~D.\ 2000, Space Science Reviews, 92, 133

\bibitem[Miyamoto \& Nagai(1975)]{1975PASJ...27..533M} Miyamoto, M.,
\& Nagai, R.\ 1975, \pasj, 27, 533

\bibitem[\protect\citeauthoryear{Morbidelli \& Levison}{2004}]{ML04}
Morbidelli, A.; Levison, H. 2004, AJ, 128, 2564

\bibitem[\protect\citeauthoryear{Pfahlmut}{2006}]{PM} 
Pfahl, E., Muterspaugh, M., 2006, ApJ, 652, 1694

\bibitem[Pichardo et al.(2004)]{2004ApJ...609..144P} 
Pichardo, B., Martos, M., \& Moreno, E.\ 2004, \apj, 609, 144

\bibitem[Pichardo et al.(2003)]{2003ApJ...582..230P} 
Pichardo, B., Martos, M., Moreno, E., \& Espresate, J.\ 2003, \apj,
582, 230

\bibitem[Press et al.(1992)]{1992nrfa.book.....P}
Press, W.~H., Teukolsky, S.~A., Vetterling, W.~T., \& Flannery,
B.~P.\ 1992, Cambridge: University Press, |c1992, 2nd ed.,

\bibitem[\protect\citeauthoryear{Schneider}{2010}]{SCH09} Schneider,
J. 2010, {\it Extrasolar Planets
Encyclopedia}. http://exoplanet.eu/catalog-all.php

\bibitem[Spurzem et al.(2006)]{2006astro.ph.12757S} 
Spurzem, R., Giersz, M., Heggie, D.~C., \& Lin, D.~N.~C.\ 2009,
ApJ, 697, 458

\bibitem[\protect\citeauthoryear{Stern}{2003}]{S} 
Stern, A., 2003, Nature, vol.242

\bibitem[\protect\citeauthoryear{Udry \& Santos}{2007}]{U07} Udry, S.,
Santos, N.C. 2007, Ann. Rev. Astron. Astrophys, 45, 397

\bibitem[Wadhwa et al.(2007)]{2007prpl.conf..835W} 
Wadhwa, M., Amelin, Y., Davis, A.~M., Lugmair, G.~W., Meyer, B.,
Gounelle, M., \& Desch, S.~J.\ 2007, Protostars and Planets V, 835

\bibitem[\protect\citeauthoryear{Weissman}{1996}]{W96}
Weissman, P. 1996, Earth Moon Planets, 72, 25




\end{thebibliography}
\end{document}